\documentclass{article}

\usepackage{arxiv}
\usepackage{authblk}
\usepackage{enumitem}
\usepackage{verbatim}
\usepackage[utf8]{inputenc} 
\usepackage{hyperref}       
\usepackage{graphicx}
\usepackage{multicol}
\usepackage[most]{tcolorbox}
\usepackage[mathscr]{euscript}
\usepackage{tikz}
\usepackage{framed}
\usepackage{xcolor}
\usepackage{setspace}
\usepackage{rotating}
\usetikzlibrary{arrows.meta, positioning, shapes.geometric}
\usepackage{multibib}
\newcites{supplementary}{Supplementary-Literature}%

\title{Design for a Digital Twin in Clinical Patient Care}

\author[1,*]{Anna-Katharina Nitschke}
\author[1,*]{Carlos Brandl}
\author[1]{Fabian Egersdörfer}
\author[2,3]{Magdalena Görtz}
\author[3]{Markus Hohenfellner}
\author[1,$\dagger$]{Matthias Weidemüller}
\affil[1]{Physikalisches Institut \protect, Universität Heidelberg \protect }
\affil[2]{Junior Clinical Cooperation Unit 
`Multiparametric Methods for Early Detection of Prostate Cancer', German Cancer Research Center (DKFZ), Heidelberg, Germany.}
\affil[3]{Department of Urology, Heidelberg University Hospital, Heidelberg, Germany.}
\affil[*]{These authors contributed equally to this work.}
\affil[$\dagger$]{corresponding author, Email: weidemueller@uni-heidelberg.de}

\begin{document}
\twocolumn[
\begin{@twocolumnfalse}
\maketitle
\begin{abstract}
Digital Twins hold great potential to personalize clinical patient care, provided the concept is translated to meet specific requirements emerging from established clinical workflows.
We present a general and unspecialized Digital Twin design combining knowledge graphs and ensemble learning to reflect the entire patient’s clinical journey and assist clinicians in their decision-making.
Such a design is predictive, modular, evolving, informed, interpretable and explainable, thus opening broad clinical applications.

\end{abstract}
\end{@twocolumnfalse}
]

\section{Introduction}
In the era of precision medicine, Digital Twins (DTs) are emerging as a long-term goal with the potential to revolutionize healthcare delivery and have experienced a surge of publications in recent years \cite{topolHighperformanceMedicineConvergence2019} \cite{katsoulakisDigitalTwinsHealth2024}.
Precision medicine aims at leveraging multi-omic, demographic, environmental, and lifestyle patient data to improve the prediction of disease occurrences and deliver personalized treatment recommendations \cite{maceachernMachineLearningPrecision2021}.
The DT as a detailed, digital replica of an individual patient extends the definition of precision medicine by enabling drug discovery and development, real-time health monitoring, and surgery planning and rehearsal \cite{ kamelboulosDigitalTwinsPersonalised2021, sunDigitalTwinHealthcare2023}.
For explicit clinical applications, DTs of organs, like the heart, have been proposed. Those are usually driven by mechanistic models, simulating variables like blood flow and blood pressure to create synthetic physiological data like photoplethysmograms \cite{mazumderSyntheticPPGGeneration2019}. Recent advances in machine learning have enabled well-performing models for many clinical tasks, which can form a basis for a predictive DT. 
The combination of mechanistic models with machine learning models is conceptually desired, but the integration of (multiple) machine learning and mechanistic models into a unified framework is still part of active research \cite{hernandez-boussardDigitalTwinsPredictive2021, corral-aceroDigitalTwinEnable2020}.

A desired application for DTs in the field of personalized medicine is the representation of patients along their whole clinical journey. Commonly, DTs for personalized medicine are tailored to a specific disease or medical question (like examples summarized within Vallé's viewpoint \cite{Valle2024}). It is difficult to extend such DTs towards other diseases or new procedures.
First approaches on how to cover clinical patient journeys by DT of patients have been presented, for example, by Voigt et al. for personalized multiple sclerosis care by outlining patient data streams, clinical workflows and procedures, existing models, prior knowledge, and open challenges\cite{voigtDigitalTwinsMultiple2021}. One central problem for the development of DTs in precision medicine is the lack of uniform methods, standards and norms. There are different challenges associated with different stages of the development of a Digital Twin, including data collection, data storage, governance, algorithmic designs and human interaction \cite{voigtDigitalTwinsMultiple2021}.

In the following, we focus on the derivation of a uniform design of a DT on the algorithmic level. De Domenico et al. argue that the complexity of disease and therefore personalized patient care require modular, non-specialized DT designs which can effectively integrate multiple subsystems for specialized cases \cite{dedomenicoChallengesOpportunitiesDigital2025}. One example of an unspecialized modular DT design has been published by Gaebel et al., using a knowledge graph-based framework \cite{gaebelDigitalTwinModular2021}.
Such knowledge graph-based frameworks can describe and connect specialized data processing models and their causal relationships, e.g., by Resource Description Frameworks (RDFs) \cite{gaebelDigitalTwinModular2021}. 
One disease-specific implementation of a knowledge graph-based DT was performed by Grieb et al. \cite{griebDigitalTwinModel2023} for multiple myeloma treatment. They implemented a recursive model structure and described implementation strategies on a technical level. They highlighted the benefits of using knowledge graphs, such as the integration of multiple heterogeneous data sources and specialised modules, as well as the inclusion of medical knowledge and guidelines. However, Gaebel et al. acknowledge that one difficulty of their design lies within the lack of a strategy for managing redundancies, which emerge from the fragmented clinical domains, where many models might provide estimates of the same information \cite{gaebelDigitalTwinModular2021}.

This perspective article presents a general and unspecialized design for a patient DT to reflect the evolving clinical patient journey based on a bipartite knowledge graph, combining patient attributes derived from patient data and preexisting predictive models. 
We designed an overhead algorithm to enable a meaningful simulation of the patient journey. The algorithm orchestrates the execution of several preexisting specialized models with similar tasks, as well as with different tasks. 
The central part of the DT's model orchestration is the introduction of fusion models that are based on ensemble learning strategies to synergize diverse model outputs into one, and a propagation and aggregation scheme that orchestrates the information flow within the knowledge graph.
Our Digital Twin design is built upon existing clinical infrastructures across departments, such as the hospital information system (HIS). Within this environment, we define a set of requirements that need to be fulfilled in order to ensure clinical acceptance and broad applicability and, hence, guide the algorithmic design.
The design bears five characterizing features. It is modular, informed, predictive, evolving, explainable and interpretable.
To enhance clarity, we illustrate specific features using two case studies, prostate cancer diagnosis and glioma treatment, throughout the article. A detailed description of those case studies is given in the supplementary material.

\section{Fundamental Considerations}
\label{Sec:DTChallenges}
\subsection{Requirements for a Digital Twin in Clinical Patient Care}
\begin{figure*}[hbt!]
    \centering
    \includegraphics[width = \textwidth]{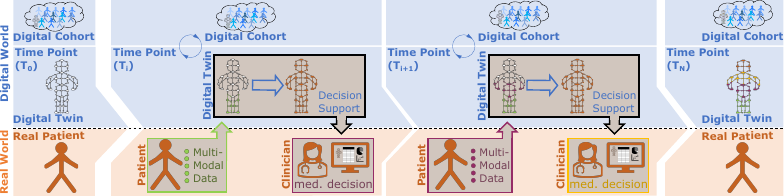}
    \caption{Visualisation of the general design of a DT for clinical patient care, generating an interface between the real world (orange; including the patient and the clinician) and the digital world (blue; including the Digital Cohort - Data Cohort and the DT - patient-specific information and algorithmic structure). Multi-modal data gets transferred from the real patient to the DT, which presents its decision support to the clinician for a medical decision.}
    \label{fig:humanDT}
\end{figure*}

The widespread adoption and efficacy of personalized models, including DT models, in clinical practice face several persistent challenges that can be categorized into three types of requirements.
The first category includes the specific requirements posed on a Digital Twin in clinical patient care, such as the holistic representation and simulation of the real world patient journey.
Secondly, we can summarize the design requirements necessary for the correct handling of data, dealing with aspects such as data missingness, data inconsistency, multi-modal data types, large datasets \cite{katsoulakisDigitalTwinsHealth2024}. The third category includes requirements related to the clinical acceptance of such a DT system, such as the adaptation to new procedures, robustness, incorporation of existing evidence, continuous learning and interpretability, especially since many machine learning models function as black boxes \cite{hernandez-boussardDigitalTwinsPredictive2021, thangarajCardiovascularCareDigital2024}.
A more comprehensive overview of the requirements of all three categories for a DT in the clinical context can be found in Supplement A.

Based upon the conceptual requirements from the first category, the characteristics of a DT for clinical patient care are visualized within Figure \ref{fig:humanDT}.
Key characteristics and definitions for any DT include the existence of a physical counterpart (one-to-one), the ability to provide a holistic representation, support for bi-directional communication, and the capability for immediate responses and operation upon data entries (real-time data) \cite{vandervalkTaxonomyDigitalTwins2020}. In clinical routine, data entries and updates are performed at clinically meaningful intervals (e.g., sub-seconds in surgery, hours in an intensive care unit, or weekly in outpatient settings). At each of these time points  $T_i$, a bidirectional communication between the real and the digital world is initiated, by passing multimodal data from the patient to the DT, expanding its patient-specific representation and allowing for predictions or simulations leading to an interpretable Decision Support output. 

This information is forwarded to the clinician in the real world to execute a medical decision related to medical interventions that lead to changes in the patient's state and/or new data acquisition. 
The collected patient information is included in the Digital Cohort, which is forming the knowledge base for any prediction model within the DT, allowing for continuous learning and extension of knowledge. Additionally, at every time point defined as the update and prediction of the DT, the current representation of the patient state is stored, and newly available data from the Digital Cohort can be used to improve and retrain the DT. The interaction between the Digital Twin and the Digital Cohort can be seen as a second type of bidirectional communication.
From a clinical perspective, it is important to emphasize that the DT system will support and supplement current clinical practice instead of replacing rigorously derived clinical knowledge and guidelines.

\subsection{Phases of Clinical Patient Journey}
A detailed understanding of the specific contexts within the clinical patient journey in which the DT will operate is essential for developing an algorithmic design that performs reliably across all relevant settings.
The explicit structure for each patient journey can differ depending on the disease type, clinical setting, and country.
Therefore, an abstract representation of the individual steps within a clinical patient journey and the related decision-making process is needed.
We derived the classification of three phases within the clinical patient journey, leading to distinct prediction tasks for the DT.

First of all, there is the \textit{observational phase}, in which complementary properties/ attributes of the patient need to be predicted. This can be viewed as a time prediction of the knowledge of the current patient state. The DT can predict the outcome of a measurement based on several models and for which the measurement procedure ideally will not change the current patient state. Two examples for such measurements are provided in the supplementary material C.1. Those are diagnostic imaging procedures and biopsies for cancer staging, each of which provides information that guides treatment decisions.

Secondly, there exists an \textit{active phase} for which the DT needs to predict changes in the patient state related to the performance of a certain intervention. This phase comprises the treatment planning, for which clinicians should be provided with a patient-specific outcome prediction and a risk profile for each potential treatment. The DT’s predictive power and reliability in that phase must be able to be evaluated by clinical endpoints (e.g., survival, recurrence, hospitalization rate). For example, as outlined in the case study of glioblastoma in the supplementary material C.2, the DT can forecast treatment outcomes given a potential chemotherapy or radiotherapy by predicting the overall survival of the patient, effectively simulating what-if scenarios.

Finally, there is a \textit{monitoring phase}, which entails making predictions regarding changes in the patient's state without any interventions.
This can be, for example, the observation of disease-specific parameters after treatment, indicating the recurrence of a disease. The DT continuously tracks, for example, follow-up biomarkers, imaging data, and patient-reported outcomes. Real-time updates of the DT may trigger early detection of tumor recurrence. In the case of a glioma, the treatment follow-up consists of regular MRI scans indicating the recurrence or progression state of the tumor.

\section{Design of the Digital Twin}
\begin{figure*}[hbt!]
    \centering
    \includegraphics[width = \textwidth]{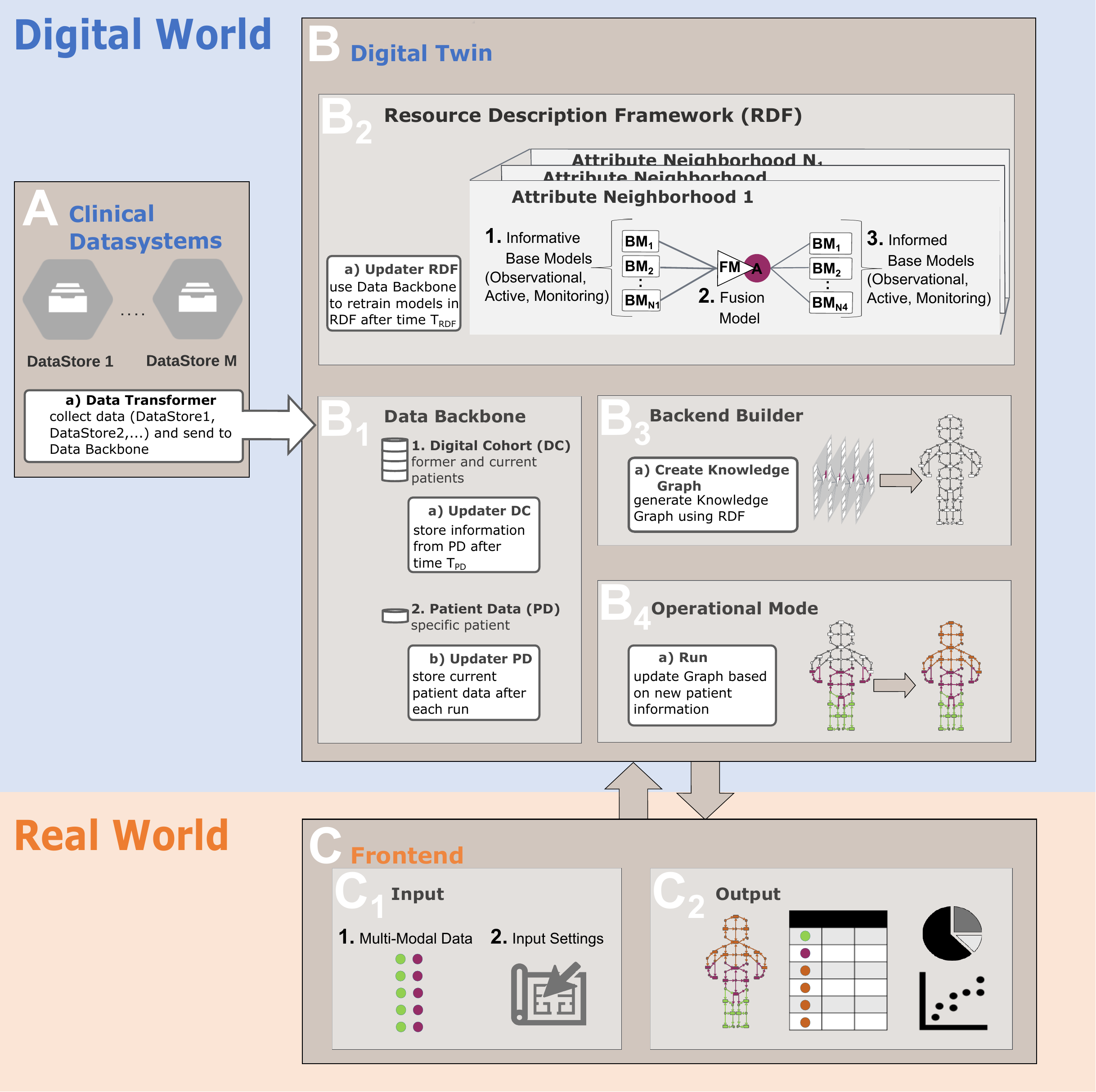}
    \caption{Schematic overview of our proposed software design for patient-centered DTs in Medicine. It consists of 3 different containers, which are labeled A-C.
    Those represent the different interacting building blocks, where A is the existing clinical data systems (hospital information system) feeding into B, the DT itself, explicitly the \textit{data backbone} (B$_1$).  The \textit{Resource Description Framework} (B$_2$) stores all available information about models and their links with attributes, which the \textit{back-end builder} (B$_3$) uses to construct a knowledge graph upon which the \textit{operational mode} (B$_4$) is going to perform predictions. The user interacts over C the \textit{frontend}, where block C$_1$ represents the inputs and C$_2$ the corresponding outputs from executing the function \textit{run} (within B$_4$). More detailed information is given in the text.}
    \label{fig:SoftwareArchitecture}
\end{figure*}
\subsection{General Design Considerations}
Figure \ref{fig:SoftwareArchitecture} schematically depicts our unspecialized DT design for clinical patient care, operating as an overhead to several pre-existing specialized models in a given clinical setting. 
\textit{Clinical datasystems} (A) include patient data that is collected across several departments and typically accessed via a hospital information system (HIS). To make the patient data accessible to machine learning algorithms, a \textit{data transformer} (A.a) needs to bring the data in a consistent format, as this is commonly collected within and distributed across diverse data formats. The \textit{data transformer} stores the structured patient data in the DT's building block called \textit{data backbone} (B$_1$). 

The overall Digital Twin design (B) consists of four components: the \textit{data backbone} (B$_1$), the \textit{resource description framework} (B$_2$), the \textit{backend builder} (B$_3$) and the \textit{operational mode} (B$_4$).
The \textit{frontend} (C) is presenting the input to the Digital Twin, consisting of the \textit{multi-modal data input} (C$_1$.1) and \textit{user settings} (C$_1$.2), as well as the DT's \textit{output} (C$_2$), visualising the final state of the knowledge graph-based patient representation. The data modalities can include laboratory tests, vital signs, medications/procedures, images, waveforms, free text, patient-reported data, devices/wearables, genomics, etc.

Specific examples of this design are provided in Supplement C, for the case of a prostate cancer biopsy (C.1) and glioma treatment (C.2).

\subsection{Data Backbone}
The first building block of the Digital Twin in Fig. \ref{fig:SoftwareArchitecture} is the \textit{data backbone} (B$_1$), which consists of the \textit{Digital Cohort} (DC) (B$_1$.1), an internal storage of the data of previous and current DTs, and the \textit{patient data} (B$_1$.2), an internal storage of all measured as well as predicted states of the current patient of interest. 
Those two are connected via an \textit{updater} (B$_1$.a), which at time intervals T$_{DC}$ passes data from the patient of interest to the DC. 
The \textit{data backbone} is the foundation on which models within the \textit{Resource Description Framework} (RDF) (B$_2$) are going to be trained via the \textit{updater RDF} (B$_2$.a) after some time period T$_{RDF}$. 

\subsection{Resource Description Framework and Backend Builder}
As visualized in Fig. \ref{fig:SoftwareArchitecture}, the \textit{Resource Description Framework} RDF (B$_2$) stores all available information to construct the knowledge graph-based patient representation. A knowledge graph is able to integrate various bits of knowledge (nodes) and their complex interrelations (edges). Two types of nodes exist within our knowledge graph design. The first node type are attributes representing patient data (e.g., biomarker, screening image, genomic mutation, clinical decisions) with their corresponding fusion model. The second node type are base models capable of predicting one set of attributes from another, including computer-interpretable-guidelines (CIG), mechanistic models and AI driven models. The included base models are mainly support systems, able to solve or predict one highly specific task. In the prostate cancer case study (C.1), such models predict biopsy outcomes, such as high grade cancer, based on distinct diagnostic features. For glioma treatment response estimation (C.2), base models perform a survival prediction. The directed edges incident at one attribute can be categorised into two types: “informed by” and “informs”.
The RDF can be organized into \textit{attribute neighborhoods} listing all base models that inform (B$_2$.1) or are informed (B$_2$.3) by an attribute. The \textit{attribute neighborhoods} have a fully modular structure and form the building blocks used by the \textit{backend builder} (B$_3$) to create a bipartite, patient-specific \textit{knowledge graph} (B$_3$.a) [$\rightarrow$ Modular Digital Twin].

Importantly, our approach is entirely agnostic towards the inner workings of any integrated model, allowing for versatile sources of information, from powerful machine learning models or simulations over computer-interpretable-guidelines (CIG) to even human experts. 
By including literature and practice-based evidence in the form of base models, we are using Informed Machine Learning within our DT design [$\rightarrow$ Informed Digital Twin].

Even if there is no difference in how the models are integrated into the knowledge graph, we can make a conceptual distinction between three prediction tasks in the different clinical phases respectively related to \textit{observational models}, \textit{active models} and \textit{monitoring models}, which will be further discussed in the Supplement B.1 [$\rightarrow$ Predictive Digital Twin]. 

As evident from the structure of the \textit{attribute neighborhoods}, several base models can predict one attribute(B$_2$.1), leading to redundancy. 
To address this, we pair each attribute with a unique model, called the fusion model (B$_2$.2) according to ensemble learning terminology. It is tasked with aggregating different statements into one consensus value.
The simplest case of two proposals being aggregated into one is illustrated in Figure \ref{fig:signature_prop} and forms the fundamental building block of the network. The final attribute prediction can then itself inform other base models (B$_2$.3). 

The fusion model (B$_2$.2) is the central mechanism that extends the Digital Twin’s capabilities beyond those of the underlying base models. Data fusion is well established as a means to improve predictive performance, even when implemented through relatively simple techniques such as weighted averaging of inputs (aggregation mode). Beyond enhancing accuracy, a carefully designed fusion model increases robustness: it can mitigate the impact of implausible or inconsistent inputs by applying robust estimators, predefined plausibility ranges, and clinician-specific weighting that reflects trust in individual data sources. Most simple non-parametric approaches for fusion models, like, e.g., weighted averages for continuous parameters or majority votes for discrete decisions, are genuinely resilient to missing inputs. They can generate outputs from any available data while progressively refining predictions as additional inputs become available. Such incremental integration further enables cross-checking for internal consistency. Failure modes, such as irreconcilable conflicts in binary decisions, become progressively unlikely as the complexity and the number of input decision grows. In the remaining undecidable situations, the fusion model needs to be able to pass such conflicting outputs back to the doctor or at least refuse to propagate the results. 

In an exploratory study \cite{BrandlIEE}, we investigated different weighting schemes for the fusion model. The base models were given either by different data modalities, pre-existing models, or different diagnostic procedures. The fusion methods included accuracy weighting, entropy weighting, linear regression, logistic regression and neural network models. For the use cases of heart disease detection and glioma diagnosis, there was no significant performance difference between those different weighting methods \cite{BrandlIEE}.

Further, we illustrate the behaviour of the fusion model using two clinical case studies that represent different fusion tasks, as shown in Supplement C. These examples range from simple overwriting modes—based on clinicians’ primacy votes—to more complex and comprehensive aggregation across different timescales of therapy options.

\subsection{Operational Mode}

Attributes determined by real world diagnostics are used as starting points, activating all available pathways to reveal the most plausible completion of all relevant attributes of the knowledge graph. This update and execution is performed via the \textit{run}-function of \textit{operational mode}'s (B$_4$) as shown in Figure \ref{fig:SoftwareArchitecture}). 
The new input is propagated through the knowledge graph until a maximally holistic attribute estimation is reached. The prostate cancer diagnosis case study in supplement C.1 illustrates such an update, using an MRI examination as example. Before the MRI, only one base model could predict high grade cancer. Afterwards, automatic evaluation of the MRI allows for three base models to predict high grade cancer, resulting in a much more profound prediction.
This state of the current best knowledge of the patient of interest is stored within the \textit{patient data} by the \textit{updater PD} (B$_1$.b). This feedback can also be used during the RDF update to adapt or refine the models, constantly improving the DT. Additionally, new \textit{base models} can be added to the RDF, and evidence-based models like CIG can be updated, leading to a constant improvement of the DT [$\rightarrow$ Evolving Digital Twin]. The computational demands will linearly grow with the number of base models and the frequency of their reevaluation, i.e., the frequency of new external inputs. The computational overhead of the fusion is small.

The propagation algorithm's design, as described in Figure \ref{fig:MainLoop}, is localized and asynchronous, meaning individual modules can perform their tasks independently. Base and fusion models constantly listen for updates on their inputs received from other models via appropriate transmission protocols. The base models will evaluate and propagate the results if an update is detected and all requirements for model evaluation are met, i.e., all input attributes are available. The propagated outputs are then registered as updated inputs by all downstream models, potentially triggering their evaluation next. This process continues until all reachable branches of the network have been activated and no more changing attributes occur. In terms of computational demands, the asynchronous implementation also means that lighter models are not slowed done by more resource-intensive simulations. Similarly, parts of the network with incomplete inputs do not prevent the evaluation of parts with complete ones.

To keep track of the origin of information, each model adds a unique signature to a provenance chain $\mathscr{P}$ that is passed along with the propagated attribute values. In effect, this list contains all attributes and models that have influenced the current value of an attribute. This serves two purposes. Firstly, it is important to maintain a degree of interpretability. Secondly, it addresses feedback loops, which describe an output that is returned to the input and are one of the key features in any complex network. 
These occur naturally when complementary sets of attributes are connected reciprocally via different models. 
Loops are problematic, as minor inconsistencies can amplify, rendering the outputs useless. The provenance chain offers a safeguard against this kind of failure. 

Based on this provenance chain $\mathscr{P}$, the fusion models' acceptance and propagation of new information are conditioned on a series of checks. Firstly, as shown in the right flow diagram in Fig. \ref{fig:MainLoop}, if the attribute is externally set, the fusion model will obviously not accept any proposals from upstream Models as it already knows the true value (overwrite mode). It will forward this value with only its signature, marking the external attribute as an origin of information. Further, if a fusion model registers its signature on the provenance chain passed along one of its inputs, it will refuse to acknowledge the update, stopping feedback loops and ensuring the algorithm terminates. 

\begin{figure}[hbt!]
    \centering
    \begin{tikzpicture}[
  node distance=0.5cm and 0.5cm,
  doc/.style={rectangle, draw, minimum width=0.3cm, minimum height=0.45cm, fill=white, inner sep=0.1cm , font=\bfseries},
  process/.style={rectangle, thick, rounded corners, draw, fill=green!20, minimum width=1cm, minimum height=0.5cm},
  func/.style={regular polygon, thick, regular polygon sides=3, draw, fill=yellow, minimum size=1.7cm, shape border rotate=270},
  decision/.style={rectangle, thick, rounded corners, draw, fill=white, minimum width=1cm, minimum height=0.5cm, text centered},
  extatt/.style={circle, thick, fill=black, minimum size=0.5cm, inner sep=0pt},
  fusion/.style={regular polygon, thick, regular polygon sides=3, draw, fill=blue!30, minimum size=0.5cm, shape border rotate=270},
  intatt/.style={circle, thick, fill=white, draw, minimum size=1.2cm, inner sep=0pt},
  halfcircle/.style={circle, thick, draw=#1, minimum size=0.5cm,
        path picture={\fill[#1] (path picture bounding box.north west) rectangle (path picture bounding box.south);}},
  arrow/.style={thick, ->, >=Stealth}
]

\node[extatt] (in1) {};
\node[fusion, left=0.1cm of in1, xshift=0.4cm, fill=blue!50] (fusion_in1) {};
\node[process, right=of in1, fill=green!50] (m1) {\tiny m.evaluate()};

\node[extatt, below=1.6cm of in1] (in2) {};
\node[fusion, left=0.1cm of in2, xshift=0.4cm, fill=black!50] (fusion_in2) {};
\node[process, right=of in2, fill=red!50] (m2) {\tiny m.evaluate()};

\node[intatt, right=1.6cm of m1, yshift=-0.75cm] (fatt) {};
\node[func, left=-0.6cm of fatt, fill=orange!50 ,align=right] (f) {};
\node[right=-1.4cm of f] (f_label) {\tiny f.aggregate()};

\node[decision, right=6cm of in1] (out1) {};
\node[decision, below=0.2cm of out1] (out2) {};
\node[below=0.2cm of out2] (etc) {...};

\node[doc, below=-0.1cm of in1, xshift=0.2cm, align=flush center] (d1) {\textcolor{blue!50}{\rule{0.2cm}{0.5mm}}};
\node[doc, below=-0.1cm of in2, xshift=0.2cm, align=flush center] (d2) {\textcolor{black!50}{\rule{0.2cm}{0.5mm}}};
\node[doc, below=-0.1cm of m1, xshift=0.4cm, align=flush center] (d3) {\textcolor{blue!50}{\rule{0.2cm}{0.5mm}} \\[-0.3cm] \textcolor{green!50}{\rule{0.2cm}{0.5mm}}};
\node[doc, below=-0.1cm of m2, xshift=0.4cm, align=flush center] (d4) {\textcolor{black!50}{\rule{0.2cm}{0.5mm}} \\[-0.3cm] \textcolor{red!50}{\rule{0.2cm}{0.5mm}}};
\node[doc, below=-0.3cm of fatt, xshift=0.5cm, align=flush center] (d5) {\textcolor{blue!50}{\rule{0.2cm}{0.5mm}} \\[-0.3cm] \textcolor{green!50}{\rule{0.2cm}{0.5mm}} \\[-0.3cm] \textcolor{black!50}{\rule{0.2cm}{0.5mm}} \\[-0.3cm] \textcolor{red!50}{\rule{0.2cm}{0.5mm}} \\[-0.3cm] \textcolor{orange!50}{\rule{0.2cm}{0.5mm}}};

\draw[arrow] (in1) -- (m1);
\draw[arrow] (in2) -- (m2);
\draw[arrow] (m1) -- ++(0.8,0) |- (f);
\draw[arrow] (m2) -- ++(0.8,0) node[above, xshift=-0.65cm, yshift=0.5cm] {\tiny m.propagate()} |- (f);
\draw[arrow] (fatt) -- ++(1.5,0) node[below, xshift=-0.0cm, yshift=-0.5cm] {\tiny f.propagate()} |- (out1);
\draw[arrow] (fatt) -- ++(1.5,0) |- (out2);
\draw[arrow] (fatt) -- ++(1.5,0) |- (etc);

\node[halfcircle, below=1.2cm of f, xshift=-0.1cm] (legend_att) {};
\node[right=-0.0cm of legend_att, xshift=0.2cm] (legend_att_label) {\tiny attribute};

\node[process, below=0.2cm of legend_att, fill=white] (legend_proc) {};
\node[right=-0.0cm of legend_proc] (legend_proc_label) {\tiny base model};

\node[fusion, right=0.5cm of legend_att_label, fill=white] (legend_fus) {};
\node[right=-0.0cm of legend_fus] (legend_fus_label) {\tiny fusion model};

\node[doc, below=0.2cm of legend_fus, fill=white] (legend_doc) {};
\node[right=-0.0cm of legend_doc, xshift=0.18cm] (legend_fus_label) {\tiny provenance chain $\mathscr{P}$};

\end{tikzpicture}
    \caption{Local structure of the knowledge graph. Outputs from two different models (green and red) are passed on to the fusion model of an internal attribute (circle). Which then gets forwarded to downstream models. The signature propagation from upstream to downstream models is illustrated by colored signatures on the provenance chain. Labels indicate subroutines from the network algorithm described in Figure \ref{fig:MainLoop}.}
    \label{fig:signature_prop}
\end{figure}

\begin{figure*}
    \centering
    \begin{tikzpicture}[
    node distance=0.5cm and 1cm,
    every node/.style={font=\small},
    startstop/.style={rectangle, rounded corners, draw=black, fill=gray!20, minimum width=3cm, minimum height=1cm},
    process/.style={rectangle, draw=black, fill=gray!10, minimum width=2.5cm, minimum height=0.7cm},
    decision/.style={diamond, draw=black, fill=orange!20, minimum width=1cm, minimum height=1.5cm, aspect=2.5},
    arrow/.style={thick, ->, >=Stealth},
    arrowd/.style={dashed, thick, ->, >=Stealth},
    parallel/.style={draw=none, font=\bfseries},
    basemodel/.style={rectangle, thick, rounded corners, draw, minimum width=1cm, minimum height=0.5cm},
    extatt/.style={circle, thick, fill=white, draw, minimum size=0.8cm, inner sep=0pt},
    fusion/.style={regular polygon, thick, regular polygon sides=3, draw, minimum size=0.5cm, shape border rotate=270, inner sep=0},
    func/.style={regular polygon, thick, regular polygon sides=3, draw, fill=white, minimum size=1cm, shape border rotate=270, inner sep=0}
]
\node (anchor) {};

\node (baseCheck) [decision, below=of anchor, xshift=-4cm] {input updated?};
\node (baseEval) [process, below=of baseCheck] {$b.evaluate()$};
\node (baseSign) [process, below=of baseEval] {$\mathscr{P} \gets \mathscr{P} \cup \{b\}$};
\node (baseProp) [process, below=of baseSign] {$b.propagate()$};

\node (fusionCheck) [decision, below=of anchor, xshift=1cm] {input updated?};
\node (isExternal) [decision, right=of fusionCheck, align=center] {real world\\input?};

\node (loopCheck) [decision, right=of isExternal] {$f \in \mathscr{P}$ ?};
\node (agg) [process, below=of loopCheck] {$f.aggregate()$};
\node (fusionSign) [process, below=of agg] {$ \mathscr{P} \gets \mathscr{P} \cup \{f\}$};
\node (fusionProp) [process, below=of fusionSign] {$f.propagate()$};

\node (extSet) [process, left=of fusionSign, xshift=-0.635cm] {$ \mathscr{P} \gets \{f\}$};

\draw [arrow] (baseCheck) -- node[left] {Yes} (baseEval);
\draw [arrow] (baseEval) -- (baseSign);
\draw [arrow] (baseSign) -- (baseProp);
\draw [arrow] (baseCheck.east) -- ++(0.2,0) node[right] {No} -- ++(0,1) -- ++(-1.5,0) -- (baseCheck.north);
\draw [arrow] (baseProp.west) -- ++(-0.5,0) node[above] {} -- ++(0,5) -- ++(+1.5,0) -- (baseCheck.north);

\draw [arrow] (fusionCheck) -- node[above] {Yes} (isExternal);
\draw [arrow] (fusionCheck.west) -- ++(-0.2,0) node[left] {No} -- ++(0,1) -- ++(1.5,0) -- (fusionCheck.north);

\draw [arrow] (isExternal) -- node[left] {Yes} (extSet);
\draw [arrow] (extSet.south) -- ++(0,-0.25) -- ++(3,0) -- (fusionProp.north);

\draw [arrow] (isExternal) -- node[above] {No} (loopCheck);
\draw [arrow] (loopCheck) -- node[right] {No} (agg);
\draw [arrow] (agg) -- (fusionSign);
\draw [arrow] (fusionSign) -- (fusionProp);
\draw [arrow] (fusionProp.west) -| (fusionCheck.south);
\draw [arrow] (loopCheck.north) -- ++(0,+0.25) node[right] {Yes}-- ++(.-7.2,0)  -- (fusionCheck.north);

\node (LabelBase) [parallel, above=0.3cm of baseCheck] {base model $b$};
\node (LabelFusion) [parallel, above=0.3cm of fusionCheck] {fusion model $f$};
\node (Viereck) [basemodel, above=of LabelBase] {$b$};
\node (kreis) [extatt, right=of Viereck, xshift=3.2cm] {};
\node (Dreiceck) [func, left=of kreis, xshift=1.4cm] {$f$};
\node (dummy) [left=of Dreiceck, yshift=0.8cm, xshift=0.1cm] {};
\node (dummy2) [right=of kreis, xshift=0.1cm] {};
\draw [arrow] (Viereck) -- ++(+2,0) node[above] {$propagate()$} -- (Dreiceck);
\draw [arrowd] (dummy) |- ([shift={(-0.25,0.25cm)}]Dreiceck.center);
\draw [arrowd] (kreis) -- (dummy2);

\end{tikzpicture}
    \caption{Flowchart illustration of the main network propagation and aggregation scheme in the operational mode. Behaviour differs between base models (left) and fusion models (right). The fusion models only propagate if their respective attribute is set externally or the provenance chain $\mathscr{P}$ does not already contain its own signature. Loops can run locally and independently of each other.}
    \label{fig:MainLoop}
\end{figure*}

Finally, the \textit{output} (C$_2$) of the Run-function is visualized in a user-oriented, intuitive way, enabling insights into predicted attribute values, algorithmic prediction, and performance measures, as well as model-agnostic interpretation methods to allow medical explainability by clinicians [$\rightarrow$ Explainable Digital Twin]. Additionally, the graph structure itself is intrinsically interpretable as a type of technical interpretability [$\rightarrow$ Interpretable Digital Twin], as the propagated signatures can be used to backtrack the influences of different attributes, as shown in the Supplement B and by portraying the graph for both clinical cases in Supplement C.1 \& C.2.

\section{Conclusion}
\subsection{Clinical Translation}
In order to translate this abstract concept into a clinical application, several aspects have to be considered.
First, validation is essential. Validation can begin with retrospective data, where the predictions of the DT are compared with ground-truth clinical outcomes. Prospective trials or real world prospective observational studies can then measure whether the DT's recommendations improve decision-making efficiency, reduce errors, or translate into better patient outcomes (e.g., survival, quality of life). Metrics such as ROC AUC, F1 scores, calibration curves or decision curve analysis can be used. 

Second, as medical decision support software, DTs must follow the guidelines of bodies such as the US FDA, the European Medicines Agency, or similar. These guidelines often require transparency regarding model updates, risk analysis and performance metrics. Implementing a robust control system for each model and systematically reporting to regulators will be critical for compliance and transparency.
Full compliance with regulatory frameworks, such as GDPR, HIPAA, or Medical Device Regulation, inevitably puts restrictions on specific parts of our design. For future implementations of our design, these regulations have to be properly taken into consideration and may be further refined in the future with increasing availability of Digital Twins

Third, operability in existing clinical infrastructures is essential. Our approach addresses this aspect by constructing a clinic-specific knowledge graph representation of the patient in accordance with clinical workflows, existing information streams and regulatory aspects.

Fourth, clinical adoption depends on clinicians' trust and understanding of the DT results. Therefore, we strongly recommend an interactive dashboard for the visualization of the DT design's intrinsic interpretability, as well as model-agnostic explanation methods (as presented in Supplemental Fig.1).
On the liability side, the role of the DT is supportive, not prescriptive: the ultimate responsibility remains with the clinician, and transparency features (e.g., explanations and confidence intervals) help clinicians assess when to rely on the DT versus their own expertise.

\subsection{Compliance with Requirements}
Each feature of the presented Digital Twin design enables compliance with distinct requirements that are posed on such a decision support system in the clinical context.
A more detailed argumentation of how the Digital Twin design complies with the requirements mentioned above can be found in the Supplement B.

A Predictive Digital Twin (through the inclusion of predictive base models) enables decision support along the whole patient journey (within all clinical phases: observation, active, and monitoring). A first type of bidirectional communication between the real and digital world, as well as the immediate response and operation upon data entries (real-time data), is facilitated.

A Modular Digital Twin will allow the implementation of several properties, such as handling multimodal input data, big data, data use optimization, information loss and missing values. Additionally, it facilitates the adaptation to new procedures, as well as reliable and robust predictions.

An Evolving Digital Twin leads to continuous learning (evolution) through a second type of bidirectional communication between Digital Twin instances of individual patients and the Digital Cohort. This simplifies the update of the decision support system and the scalability to new settings.

An Informed Digital Twin that is evidence-adaptive can be generated by including computer-interpretable clinical guidelines (as a type of medical evidence) as base models. The inclusion of Informed Machine Learning using clinical guidelines for final hypothesis validation leads to a third type of bidirectional communication between experts and the Digital Twin.

An Interpretable and Explainable Digital Twin will enhance clinical acceptance by the interpretable algorithmic design of the bipartite knowledge graph, explainable visualization methods, and an interactive output design.

\subsection{Limitations}
Despite the strengths of the DT design already presented, it has design-specific limitations, like the modularity of the Digital Twin. It can be understood as a trade-off between increasing the relative amount of training data (input feature) and losing information between features by splitting them up. Furthermore, our design relies on the availability of data and specialized base models. 

Generally, for any DT, data of higher quantity and breadth needs to be collected to be able to use the DT as a holistic decision support system. A general potential risk for the introduction of biases comes from supervised machine learning, focusing on the identification of structures frequently represented within the data.  In the ideal case, hospital information systems from several clinics can be linked together, forming a stronger database to further improve personalized care. In this way, Digital Twins of clinical patients can even be enabled to cover rare clinical cases.

Overall, broad service accessibility of the DT systems needs to be ensured, as it could otherwise be a driver for inequality in health-service provision \cite{bruynseelsDigitalTwinsHealth2018} and widen a socio-economic gap \cite{popaUseDigitalTwins2021}.
Therefore, further improvement, scaling, critical analysis, and risk assessments of any DT designs are crucial.

\subsection{Future Opportunities}
Our scalable approach allows the application to several settings and, therefore, can be applied to a broad variety of medical cases, as exemplified for two cases in Supplement C ("diagnosis of prostate cancer" and "survival prediction in glioblastoma"). 
Another example is cardiology, where many mechanistic, evidence-based models are available. Our DT successfully enables the combination of mechanistic with AI-based models.
In oncology, e.g., for prostate cancer, multi-modal data is becoming increasingly relevant \cite{goldenbergNewEraArtificial2019}, and the heterogeneity of the disease leads 
to a variety of treatment options and patient journeys \cite{haffnerGenomicPhenotypicHeterogeneity2021}, which can all be represented in our graph structure. 

Beyond cardiology and oncology, DTs have broad potential to improve clinical decision support in many clinical fields, e.g., pulmonology (for managing chronic respiratory diseases like asthma by predicting exacerbations), endocrinology (for continuous monitoring and prediction of blood glucose trends to improve treatment strategies in diabetes), critical care (for real-time monitoring and early-warning systems in intensive care units), surgical planning (for creating patient-specific models to optimize preoperative planning and predict surgical outcomes) as well as infectious diseases (for modeling the spread of infections, which could aid in personalized treatment during outbreaks). 

The predictive nature of the DT over several steps in the patient journey can also improve early diagnosis or prevention of diseases \cite{sunDigitalTwinHealthcare2023}. Due to the nature of limited clinical data before a disease, the DT will need large-scale routine medical data with socio-economic and behavioural data. In addition, digital devices, such as wearables that track physiological parameters, can enable a new dimension of predictive modelling. First applications might be common chronic diseases like diabetes or hypertension.

Therefore, given an appropriate environment with the availability of data and base models, our DT can support clinicians and doctors in the prevention, diagnosis, staging, treatment and follow-up of the patient. To improve the acceptability and also to simplify the validation, starting with small building blocks is favourable. This can be a single decision support tool based upon multiple base models or a tool using the combination of two base models through an intermediate attribute. Due to its adaptability, an extension can easily be added later on. The tasks the digital twin can handle will grow with time. At the same time, the interface will stay similar, making adoption easier.
We have developed and analyzed a first building block, i.e., for the fusion of CIGs with ML models for glioma classification \cite{BrandlIEE} and in the case of prostate cancer for diagnosis in order to minimize biopsies.

\section*{Acknowledgements}
This work was realized through support by the German Federal Ministry for Economic Affairs and Climate Action (funding \#01MT21004F - CLINIC5.1).\\
This work is funded by the Deutsche Forschungsgemeinschaft (DFG, German Research Foundation) under Germany’s Excellence Strategy EXC2181/1-390900948 (the Heidelberg STRUCTURES Excellence Cluster) and
funded by the Federal Ministry of Education and Research (BMBF) and the Ministry of Science, Research and Arts of Baden-Württemberg as part of the Excellence Strategy of the federal and state governments (Field of Focus - University of Heidelberg).\\
The funders played no role in study design, concept development, or the writing of this manuscript. 

We want to thank all the project partners of the CLINIC5.1 consortium, in particular Stefan Duensing, Annette Duensing, David Bonekamp, Clara Meinzer, Matthias Rath, and Martina Heller for the numerous insights into oncological patient diagnosis and treatment.  
We also thank Ullrich Köthe, Pingchuan Ma, and Björn Ommer for the fruitful discussions of our concept's technical side. \\
\section*{Author contributions}
All authors: Development of the general concept.
A.N and C.B.: Algorithmic and software development of the concept,  Writing—original draft, Writing—review and editing.
A.N and F.E.: Visualizations.
F.E.: Algorithmic development, Writing—original draft of operational mode, Writing—review and editing.
M.G.: Clinical insights, Writing-original draft of clinical translation, Writing—review and editing.
M.H.: Clinical insights, Writing—review, Funding acquisition.
M.W.:  Writing—review and editing, Funding acquisition, Supervision.
All authors have read and agreed to the submitted version of the manuscript.
\section*{Competing Interests}
The authors declare no competing interests.

\bibliographystyle{unsrt}  
\bibliography{bibliography}

\title{Supplement - Design for a Digital Twin in Clinical Patient Care}

\twocolumn[
\begin{@twocolumnfalse}

\maketitle

\end{@twocolumnfalse}
]
\appendix
\section{Requirement for Digital Twin in Clinical Patient Care}
\label{sec:Requirements}
As presented in Section 2.1 the requirements of a DT in the context of clinical care can be divided into three categories: Requirements concerning the DT itself, requirements related to data handling, and requirements associated with clinical acceptance. In the following, we identified the relevant requirements that are fulfilled by our design's five main features (Predictive, Modular, Evolving, Informed, Interpretable and Explainable) as introduced in sections 3.3 and 3.4 of the main text. The requirements are based on the discussion presented in Schwartz et al. \citesupplementary{schwartzDigitalTwinsEmerging2020} and Kelly et al. \citesupplementary{kellyKeyChallengesDelivering2019}.

\setlist[enumerate,1]{label=,leftmargin=*} 
\renewcommand{\labelenumi}{R\arabic{enumi}} 
\setlist[enumerate,2]{label=R\arabic*, ref=R\arabic*, labelwidth=3em, labelsep=0.5em, leftmargin=3em}
\begin{enumerate}
    \item \textbf{DT requirements} \hrule
    \begin{enumerate}
        \item Holistic representation of the patient journey
        \begin{enumerate}[label = R1.\arabic*]
            \item Operate in observational phase
            \item Operate in active phase
            \item Operate in monitoring phase
            \item Enable the prediction of the time evolution of parameters 
        \end{enumerate}
        \item Facilitate a bi-directional communication
        \item Responds and operates upon data entries (real-time) 
    \end{enumerate}

    \item \textbf{Data handling requirements} \hrule
    \begin{enumerate}[resume] 
        \item Multi-modal input data
        \item Deal with big data
        \item Optimize data use
        \item Avoid information loss
        \item Handle missing values 
    \end{enumerate}

    \item \textbf{Acceptance requirements} \hrule
    \begin{enumerate}[resume] 
        \item Ease adaptation to new procedures
        \item Achieve reliable and robust predictions 
        \item Continuous learning
        \item Update decision support system
        \item Ease integration into clinical workflow and scalability to new settings
        \item Informed by medical evidence 
        \item Minimize human barriers
        \begin{enumerate}[label = R15.\arabic*]
            \item Enable interpretability
            \item Facilitate explainability 
        \end{enumerate}
    \end{enumerate}
\end{enumerate}

\section{Features of the Digital Twin Architecture } 
\label{sec:Features}
In the following we specify desired features and discuss how our proposed structure is dealing with the requirements introduced in Part \ref{sec:Requirements} by advantageously combining several methodologies making up the core features of a DT that is \textbf{Predictive}, \textbf{Modular}, \textbf{Evolving}, \textbf{Informed}, \textbf{Interpretable and Explainable}.

\subsection{Predictive Digital Twin}
\begin{tcolorbox}[title=F1: Predictive,
title filled=false,
colback=black!5!white,
colframe=black!50!white]
The inclusion of predictive base models (clinical phase-specific: observational, active, monitoring) enables decision support and time evolution simulation along the whole patient journey. A first type of bidirectional communication between the real and digital world in real time is facilitated.
\end{tcolorbox}

Our Digital Twin design is the first to explicitly follow the patient journey through different medical decision time points.
The modularized design of the network algorithm facilitates the integration of diverse base models (B$_2$.1 \& B$_2$.3), covering a variety of prediction tasks. 
Similar to the model for a structural DT of an unmanned aerial vehicle presented by Kapteyn et al. \citesupplementary{kapteynProbabilisticGraphicalModel2021}, distinct time phases can be distinguished.
They defined one phase in which complementary information gets observed to better understand and describe the physical asset, called the calibration phase. In our case, this corresponds to the observational phase for which a change in the knowledge about the patient state is performed. Kapteyn et al. defined a second phase, called the operational phase, in which a fixed set of parameters is observed over time. We additionally distinguish the two model prediction tasks for which a change in the patient state can occur. One would be due to the performance of an intervention (like treatments), described as the active phase. Second, a change in the patient state due to parameter evolution (like active surveillance), described as monitoring phase.
Hence, equivalently to Kapteyn et al., depending on the current phase different methods can be used to connect the information gained over time and connect the different steps of the patient journey.\\

\textbf{R1.1: Observational Phase -}
An increasing number of parameters are observed to better understand and precisely describe the patient's health state. For example, different screening methods can be used in the diagnosis phase. After each information gain, the patient's health state is estimated, and the question is asked whether the patient shows a high probability of the need for treatment and needs to be passed to the next step within the patient journey.
The reflection of this procedure in the digital world can be understood as a form of Bayesian inference, as mentioned by Kapteyn et al. \citesupplementary{kapteynProbabilisticGraphicalModel2021}: "Using this observed data, we perform a Bayesian update on our prior estimate in order to produce the posterior estimate [...]."
At each time point, there is a prior belief about a patient's state parameters, which are adjusted according to new insights, progressively individualizing the DT.
In this phase, the models that comprise the DT structure at each time point, which include data-driven models, need to be trained on different parameter spaces and hence are independent from one another.

\textbf{R1.2: Active Phase -}
During this phase, the patient undergoes one or more interventions. A more or less consistent set of parameters is monitored over time, reflecting the patient's health status and any changes induced by interventions. By directly connecting the DT to clinical monitoring systems, this can happen in real-time. Predicted attributes will either encode recommendations for future interventions or anticipate their outcomes. Examples of such attributes include the likelihood of disease recurrence or the quality-adjusted life years, which balance the quantity and quality of the patient’s remaining life \citesupplementary{sun_prediction_2023}.
Further instances of active phase modelling are Capelli et al., who explored computational models for treatment planning in congenital heart disease \citesupplementary{capelli_patient-specific_2018}, or Smith et al., who have developed a personalized treatment planning model that integrates physical, biological, and clinical factors to optimize IMRT plans for prostate cancer \citesupplementary{smith_personalized_2016}.
By adding attributes reflecting not only the patient state but also the state of the clinical workflow, our network-based approach allows for the flexible modelling of a potentially very complex decision logic, where the performance of future interventions is conditioned on the steps taken in the past. This path dependency is often present in clinical workflows. The corresponding models must account for attributes that indicate prior interventions to achieve this.
Many existing models, however, provide only general predictions about future developments. In such cases, their outputs can be incorporated into the network like any other attribute, with the implicit understanding that these projections represent future scenarios.

\textbf{R1.3: Monitoring Phase -}
In the monitoring phase, a change in the patient state without the performance of an intervention is needed. 
In the clinical context, this is the case for monitoring in-patients, e.g., in intensive care units, as well as for out-patients, e.g., for monitoring the patient's health status after a treatment has been completed. 
For in-patients, real-time prediction based on sensory data is a typical task \citesupplementary{ShaikRemotePatientMonitoring2023}. 
Examples include the prediction of sepsis \citesupplementary{doi:10.1126/scitranslmed.aab3719}, near-time mortality prediction \citesupplementary{xiaLongShortTermMemory2019}, and prediction of cardiac arrest in emergency departments \citesupplementary{ongPredictionCardiacArrest2012}.

For outpatient care, the monitoring phase typically corresponds to the final part of the clinical patient journey after treatment has been successfully completed and the possible recurrence of the disease is of interest. Such outcome prediction can take various forms.
Especially in oncology, cox-regression models predict the probability of progression-free survival for a time period \citesupplementary{campbellOptimumToolsPredicting2017, elhajiEvolutionBreastCancer2023}. In general, clinical outcome prediction aims to detect deterioration of the patient, such as cardiac arrest, mortality, or intensive care unit (re-)admission \citesupplementary{Shamout_MachineLearningClinical2021}

Apart from data-driven prediction models, this also includes mechanistic models to simulate biological processes. For the application considered here, the most appropriate method has been presented by Masison et al. \citesupplementary{masisonModularComputationalFramework2021} as they have been developing a hub- and-spoke modular design for their simulation-based medical DT. They presented a simulation solver that could alternatively be used instead of the Fusion method and orchestrates the execution of individual submodules in the biological process. 

\textbf{R1.4: Time Evolution of Parameters -} A model task that can occur within the active and monitoring phase is the prediction of the time evolution of certain clinical parameters, crucial for further patient journey.
Models provide a continuous prediction of the temporal evolution of certain attributes and can give predictions at variable times. For these, an extra time attribute can be introduced that the clinician can manually set. The outputs from these models now effectively live in their own time zone that the propagation algorithm is not (and does not need to be) aware of. In principle, a complete copy of the network can be attached to the new time-projected attribute, which, utilizing the usual propagation rules, would project the entire patient state to the new time zone.
A variety of methods is known for clinical time series prediction, ranging from simpler statistical models like Autoregressive models, Linear Dynamical System models, and Gaussian Process models (discussed by Liu and Hauskrecht \citesupplementary{liuClinicalTimeSeries2015,liu_personalized_2017}) to complex biophysical simulations (see Chase et al. \citesupplementary{chase_next-generation_2018}).
The best-suited method for each setting depends on several factors like the time intervals, the parameter types, the type of noise, and the prediction or detection goals. Kapteyn et al. \citesupplementary{kapteynProbabilisticGraphicalModel2021} mathematically described this phase by \textit{sequential Bayesian inference}, on the output prediction of the specific models at one time point that in the end enables key capabilities of the DT. 

\textbf{R2: Bidirectional Communication -} Our presented algorithmic structure of a DT serves well for all clinical patient journey phases. Hence, a fundamental requirement, the prediction of the patient state over time, is fulfilled. Through the inclusion of clinical patient data into the Digital World and by transferring the DT prediction through the dashboard into the real world, bidirectional communication is facilitated.

\textbf{R3: Real Time Data -} An important quality of a patient DT is the ability to enable bidirectional communication in real time. In clinical routine, real-time often implies data updates at clinically meaningful intervals (e.g., sub-seconds in surgery, hours in an intensive care unit, or weekly in outpatient settings). Therefore, the link to the different clinical data systems must be established. An approach for a DT platform to establish such a connection on a software framework level has been proposed by Petrova-Antonova et al. based on different web services \citesupplementary{petrova-antonovaDigitalTwinPlatform2020}. 
With our DT design, we do not impact the information flow or data collection from the patient to the data base (hospital information system). The focus of our design is to ensure that the DT is presenting the best possible representation of the patient. Therefore, the operational mode of the DT is triggered as soon as new information about the patient becomes available.

\subsection{Modular Digital Twin}

\begin{tcolorbox}[title=F2: Modular,
title filled=false,
colback=black!5!white,
colframe=black!50!white]
 A modular decision support system will allow the implementation of several properties, such as handling of heterogeneous input/mixed-type data; big data; information loss and missing values; data use optimization; scalability /adaptation to new procedures; reliable and robust predictions; validation.

\end{tcolorbox}

As visualized in Figure 2 of the main text, the patient-specific DT is configured by the \textit{backend builder} (B$_3$). The generated bipartite knowledge graph links patient attributes and models. The \textit{attribute neighborhoods} are best described as an ensemble learning approach, which is modular by its nature. Thus, each base model (B$_2$.1 \& B$_2$.3), processes a part of the known parameter space. In a second step, a final combination of the module output is given through a fusion model (B$_2$.2) that returns an overall prediction.
In the following, the different challenges tackled by the modularity of our ansatz are further discussed.

\textbf{R4: Multimodal Input Data -  } 
Prediction tasks on multimodal input data are still active research area in Machine Learning. Medical decisions are made on a variety of data, which are of different data formats. Combining these multimodal data for medical decision support systems is not straightforward, as algorithms usually process only one class of data (text, image, numbers, etc.). By using the modular approach to combine the predictions made on different parts of the data, each base model in the \textit{attribute neighborhood} can process its corresponding data type. Current approaches to deal with mixed or multi-view data are reviewed by Li et al. \citesupplementary{liReviewMachineLearning2018}. They presented a general overview of how data fusion can be included through machine learning techniques categorised into early, intermediate, or late integration methods. In early integration methods, features from different data are concatenated into a single feature vector before fitting an unsupervised or supervised model. As this approach seems straightforward and intuitive, constructing a model that is able to deal with this kind of input vector is not easy, as further feature preprocessing might be needed. The intermediate approach involves data integration in the learning process and into the model design. The late integration method, for which separate models (base models) are first trained on the individual data subsets, involves a combination of those individual outputs to a final response and is comparable to our approach, which uses fusion methods (B.d.3). \\

\textbf{R5: Big Data - }
As more and more data is generated, the computational power for processing this data needs to be increased. One solution is to diversify the data analysis by training models only on parameter subsets of the whole data. Therefore, less computational power is needed for training the base models, as each of them only works with a subspace of the total parameters.

\textbf{R6: Data Use Optimization - }
At the same time, our approach helps optimize data use. By dividing the feature space, the relative training data available per feature is increased. Thus, the approach also works in cases where only a limited number of patients are available, like for rare diseases. Additionally, some patients only undergo specific procedures or have a high ratio of missing data, so only very limited parameters can be added to the subsets accordingly. This can further increase the amount of training data per feature subspace.

\paragraph{R7: Information Loss - }
Often, when building clinical decision support tools, not all available features are used by the developed algorithms. The features are usually selected by having prior knowledge of the clinicians or by using feature selection algorithms. But some of them could contain so far unknown meaningful information, which is why their inclusion could be advantageous. As the precise representation of the current physical condition of the patient is sought, all available information should be used in the best possible way. Hence, additional base models could be trained upon available but unused parameters and added to the fusion model. Thus, our modular approach reduces the risk of dismissing potentially valuable data.

\paragraph{R8: Missing Values - }
Our DT structure can deal with missing values, missing measurements, or measurements being executed in a different order, as the structure of the algorithm can be adjusted very easily. In the case of a predictive task based on multiple base models, missing values would lead to the corresponding base models not being executed. But due to the redundancy by other base models, it would be just the fusion model (B$_2$.2) that needs to adapt to a different number of inputs. Especially if the used fusion model is linear, this process would be resolved by a simple renormalization of the base model weight (see publication Clinical Decision Support System (CDSS) by Brandl et al.\citesupplementary{BrandlIEE}).  Usually, for other classification algorithms, a problem specific adjustment needs to be made to deal with this problem. Our approach is, therefore, easily generalizable, model-agnostic, and unspecific. 

\paragraph{R9: Adaption to New Procedures -  }
The inclusion of new decision support tools and practices is a tedious and time consuming process. The modular approach is making the implementation of a newly developed algorithm easier for CDSS developers, as they can use the existing infrastructure. As the software interface does not change, the adaptation to new procedures is simplified for clinicians. Furthermore, less implementation effort is needed. 

\paragraph{R10: Reliable and Robust Predictions - }
An important requirement for a medical support system is a reliable and robust prediction. As ensemble methods take into account several opinions and votes of different classifiers, their variance might decrease through averaging over these stories (as shown by publication Brandl et al. \citesupplementary{BrandlIEE}). Instead of constructing the best possible machine learning method, the approach is to use the wisdom of the crowds, in which independent classifiers form one decision together. The improvement by using Mixture of Experts is well described in the Literature  \citesupplementary{yukselTwentyYearsMixture2012, haykinNeuralNetworksComprehensive1999}.

 \subsection{Evolving Digital Twin}
\begin{tcolorbox}[title=F3: Evolving,
title filled=false,
colback=black!5!white,
colframe=black!50!white]
A second type of bidirectional communication between Digital Twin instances of individual patients and the Digital Cohort will lead to continuous learning (evolution).
\end{tcolorbox}

\textbf{R11: Continuous Learning -}
Bidirectional communication between individual patients' DT instances and the DT aggregate covering the full patient cohort can be realised by continuous learning/ online learning techniques.
The modules in the \textit{RDF} (B$_2$) would therefore need to be retrained after a specified time period on the updated \textit{Digital Cohort} (B$_1$.1), which itself is updated by new \textit{Patient Data} (B$_1$.2). With each iteration the data foundation for the base models (B$_2$.1\&3) and the fusion models (B$_2$.2) improves. This process of iteratively retraining the models is called online learning and is in our design controlled by the Updater RDF (B$_2$.a). The whole process can be automated after data quality is ensured. In this way, the DT can additionally adapt to slowly changing factors in the clinical patient population in a temporal manner, like an aging society, resulting in an overall higher mean age.

\textbf{R12: Updating Decision Support Systems -} Other decision support tools usually need to get updated regularly to include new characteristics found or changes in the population. The update is generally associated with repeated implementation of clinical studies, resulting in lots of work and costs. By using online learning, the algorithms used for the DT can adapt over time. This is done by regularly including new patients into the \textit{Digital Cohort} and retraining of the model.

\textbf{R13: Scalability to New Settings - } The proposed architecture is a general approach to clinical patient care, independent of an explicit clinical setting and explicit knowledge about the disease processes. The DT architecture is adapted to a specific application by choosing suitable base models and data systems. Therefore, the presented DT design is very flexible and can easily be used for a variety of medical patient journeys as well as different medical fields. Through online learning, one can further advance the algorithm to handle new populations from different countries, ethnicities, or even just different institutions. A recalibration would be generally necessary if the solution were implemented for a broader patient cohort, as otherwise, there might be the problem of sampling bias due to intrinsic and extrinsic demographic heterogeneity of the training data \citesupplementary{corral-aceroDigitalTwinEnable2020}.

\subsection{Informed Digital Twin }
\begin{tcolorbox}[title=F4: Informed,
title filled=false,
colback=black!5!white,
colframe=black!50!white]
A Digital Twin that is evidence-adaptive can be generated by including computer-interpretable clinical guidelines as base models.
The inclusion of Informed Machine Learning using clinical guidelines for final hypothesis validation leads to a third type of bidirectional communication between experts and the Digital Twin.
\end{tcolorbox}

\textbf{R14: Informed by medical evidence -} It has so far seemed as if evidence-based knowledge and knowledge extracted through machine learning from data would face each other, although the combination of both seems promising and is desired \citesupplementary{scottMachineLearningEvidenceBased2018, abujaberHarnessingMachineLearning2022}. In accordance with the discussion provided by Brandl et al. \citesupplementary{BrandlIEE}, we aim to generate a DT that includes practice-based evidence, e.g., in the form of computer-interpretable clinical guidelines (CIG).
There, we explained the possibility of using multiple base models to validate the guidelines, which can be controlled by increasing or decreasing the guidelines’ weight in the fusion model. In case of different opinions on guidelines and remaining base models, a reasoning as to where these differences come from will be helpful for the clinicians. Therefore, the interpretability and explainability (R17) of the individual models will help clinicians to make an informed decision. 

An evidence-adaptive CDSS can be generated by including CIG as a base model (B$_2$.1\&3) within the graph structure. Hence, Feature 4 "Informed" of our proposed DT design is based on Feature 2 "Modularity", but will be discussed separately due to its importance for the CDSS.
The fusion models can effectively model the degree of trust in the guidelines by increasing or decreasing the guideline’s weight, which offers a new possibility to validate different Machine Learning algorithms or evidence-based nomograms, as the fusion model calculates the weights of an algorithm in the ensemble based on its performance.
Additionally, the DT could also learn to only operate within the boundaries of the guidelines, as the result of the \textit{Operational Mode }B$_4$ can be filtered to comply with the guidelines.

Internally, the discrepancy between the prediction and the guidelines will then be evaluated post-intervention and could result in new insights into the disease. 
The adaptation of the guidelines and, therefore, the evolving state of knowledge, is easily possible for the DT. A holistic approach is formulated that tries to make the best out of all available information. 
Therefore, Informed Machine Learning by using clinical guidelines for final hypothesis validation will lead to a type of \textit{bidirectional communication between researcher/experts and Digital Twin}.

Additionally, one can determine if an algorithm performs better on a small group of patients than the total cohort, which offers a possibility to detect bias in that algorithm. 

Furthermore, the validation of the current guidelines is possible as we will explain in the next paragraph.

\subsection{Interpretable and Explainable Digital Twin}
\label{sec:InterpretableExplainable}
\begin{tcolorbox}[title=F5: Interpretable and Explainable,
title filled=false,
colback=black!5!white,
colframe=black!50!white]
The clinical acceptance of the Digital Twin will be enhanced by interpretability in the algorithmic design due to the knowledge graph structure and explainability based on visualization methods and an interactive output design.
\end{tcolorbox}

\textbf{R15.1: Interpretability -} In the medical context, algorithms' interpretability and explainability are mandatory.
Interpretability is a technical property of algorithms \citesupplementary{barredoarrietaExplainableArtificialIntelligence2020}, thus it has to be considered in the design of the Digital Twin. 
The base models and fusion models in our knowledge graph can be interpretable. 
For some examples, we have shown that more complex, non-linear fusion models do not necessarily lead to better predictive power \citesupplementary{BrandlIEE}.
The causal hierarchy between different parts of the knowledge graph can be extracted from the provenance chain passed along with all predictions. These allow the identification of used features and tracking the influence of the different base models.
Subnetworks contributing to an attribute's prediction can be visualized, and the user can be given the option to disable branches that seem to influence the attributes in question toward subjectively implausible results.

Although using a modular framework to gain insights and interpretability seems straightforward, there is not much literature on the topic. This might be due to the fact that a mixture of experts framework does not include the generation of (distinct) feature subspaces. Instead, all base models work on the same full parameter space, and the individual instances (patients) are assigned to a model. For that case, it has been shown that using only interpretable models, like Logistic Regression or (Soft) Decision Trees as Base Models and a Deep Neural Network as Assignment Network/Gating Network achieves similar performance to black-box models \citesupplementary{ismailInterpretableMixtureExperts2022}. As the models are interpretable by nature, a partially interpretable Algorithm on individual instance and feature level is constructed. However, in that approach, it is not possible to manually ex-/include individual models.

\begin{figure}[hbt!]
    \centering
    \includegraphics[width = 0.5\textwidth]{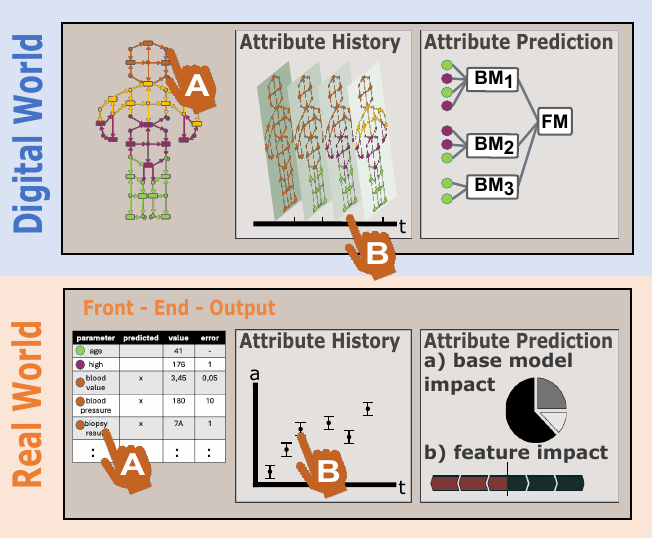}
    \caption{Visualisation of the medical Interpretability by the visualisation of the attribute history and base model impact, as well as feature impact on the attribute prediction. In the digital world, this corresponds to: the selection of a node of the knowledge graph (A), the selection of one execution of the run-function in the Operational Mode (B) and the attribute prediction via an ensemble model.}
    \label{fig:OnOffModels}
\end{figure}

\textbf{R15.2: Explainability - } Beyond interpretability, a user-based property called explainability \citesupplementary{barredoarrietaExplainableArtificialIntelligence2020} can be generated by technical tools, helping the user to understand the model, or can be developed by the user on their own in an interactive manner. 

Technical tools will be used in settings in which non-interpretable models have to be used for performance aspects. These are model-agnostic (and model-specific) explainability methods, which are well-summarised in literature \citesupplementary{barredoarrietaExplainableArtificialIntelligence2020, molnarInterpretableMachineLearning2019}.

Another way of achieving explainability is by an interactive design, allowing the user, i.e., the medical doctor, to understand the model and develop their own intuition. This can be achieved by a well-designed dashboard. The dashboard output presented in Figure \ref{fig:OnOffModels} as "Front-End-Output" shows, for example, how an attribute's prediction is impacted by the base models and the attributes through, e.g., SHAP plots \citesupplementary{lundbergUnifiedApproachInterpreting2017}.
The proposed approach, using a modular structure, introduces the ability to easily interact with the algorithm's structure, allowing the physician to develop an intuition about its decision-making process. This is represented in Figure \ref{fig:OnOffModels}, which shows an example dashboard for the clinician in the real world. The clinician is able to see which values are predicted or measured, the value and its error (A). If several values have been measured or calculated, the history of the attribute value can be visualized (B). For each time-point, the base model impact (a) and feature impact (b) on the attribute prediction can be visualized. The clinician can also directly test hypothetical scenarios (e.g., “What if the patient’s age is 10 years older?”). For clinical translation and acceptance, user testing with clinicians to refine the front-end, interpretability features, and iterative improvements is elementary. 
Figure \ref{fig:OnOffModels} additionally visualizes what this type of interaction represents on the algorithmic level: the selection of a node of the knowledge graph (A), the selection of a completed graph after the performance of the run-function in the \textit{operational mode} (B) and the attribute prediction via an ensemble model.

\section{Clinical Examples}

\begin{figure*}[h!]
    \centering
    \includegraphics[width=\textwidth]{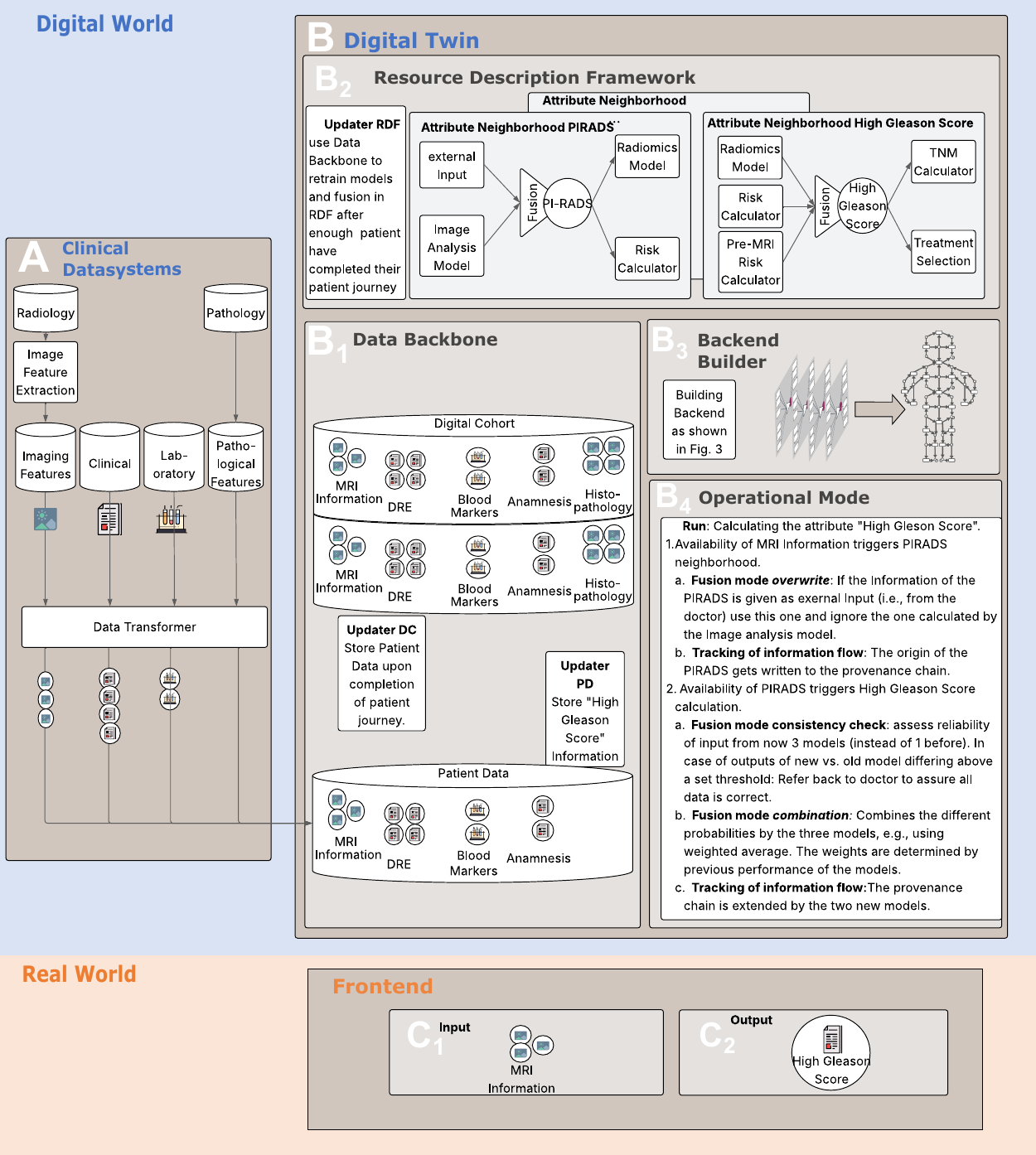}
    \caption{Schematic overview of our proposed software design for patient-centred DTs in prostate cancer diagnosis. The figure is similar to Fig. 2 of the main text, showing the Clinical Datasystems (A), the DT itself (B), and a Front-end (C). The DT consists of the \textit{data backbone} (B$_1$), the \textit{Resource Description Framework} (B$_2$), which stores all available information about models and their links with attributes, which the \textit{back-end builder} (B$_3$) uses to construct a knowledge graph upon which the \textit{operational mode} (B$_4$) is performing predictions. Structured data is visualized as small circles, with the icon depicting the original data source of the clinic. The full knowledge graph for our example is shown in Figure \ref{fig:ProstateGraph}}.
    \label{fig:ProstateFig2}
\end{figure*}

Our first example represents the observational phase of the clinical patient journey, in which information is collected from the patient via diagnostic procedures and the patient's health state can be viewed as more or less stable. The example is showcasing the process of an oncological diagnosis before biopsy, in the case of prostate cancer. As a second example, we outline a case for survival prediction of a glioblastoma patient to exemplify the application in the later stages of the clinical patient journey, representing what we have defined as the monitoring phase.

\subsection{Diagnosis of Prostate Cancer}

In our example scenario, we assume a "Patient John Smith" is receiving standard screening procedures in a clinical setting in urology, which include the performance of an anamnesis, laboratory tests of blood markers (such as PSA) and a digital rectal examination (DRE). 
As a next step, imaging using MRI is performed to identify potential lesions for an MRI-guided biopsy \citesupplementary{haringS3LeitlinieProstatakarzinom2021}. From the MRI screening, information including an overall score of the malignancy (PI-RADS), prostate volume measurement and additional radiomics features are collected. Based on the results of all screening procedures and predefined clinical guidelines, the responsible clinician needs to assess the need for the performance of such a biopsy. The aim of a biopsy is to better assess the stage and malignancy of a potential prostate tumor. Therefore, a grading called the Gleason score (GS) is performed by the pathologists from the tissue samples that are collected via the biopsy. As there are several potential complications and a biopsy is an invasive procedure, it is desirable to avoid unnecessary biopsies. Therefore, there are many risk prediction models available that aim to estimate the risk of a high GS through the analysis of available patient information \citesupplementary{vanboovenSystematicReviewArtificial2021}. 

\begin{figure*}[h!]
    \centering
    \includegraphics[width=\textwidth]{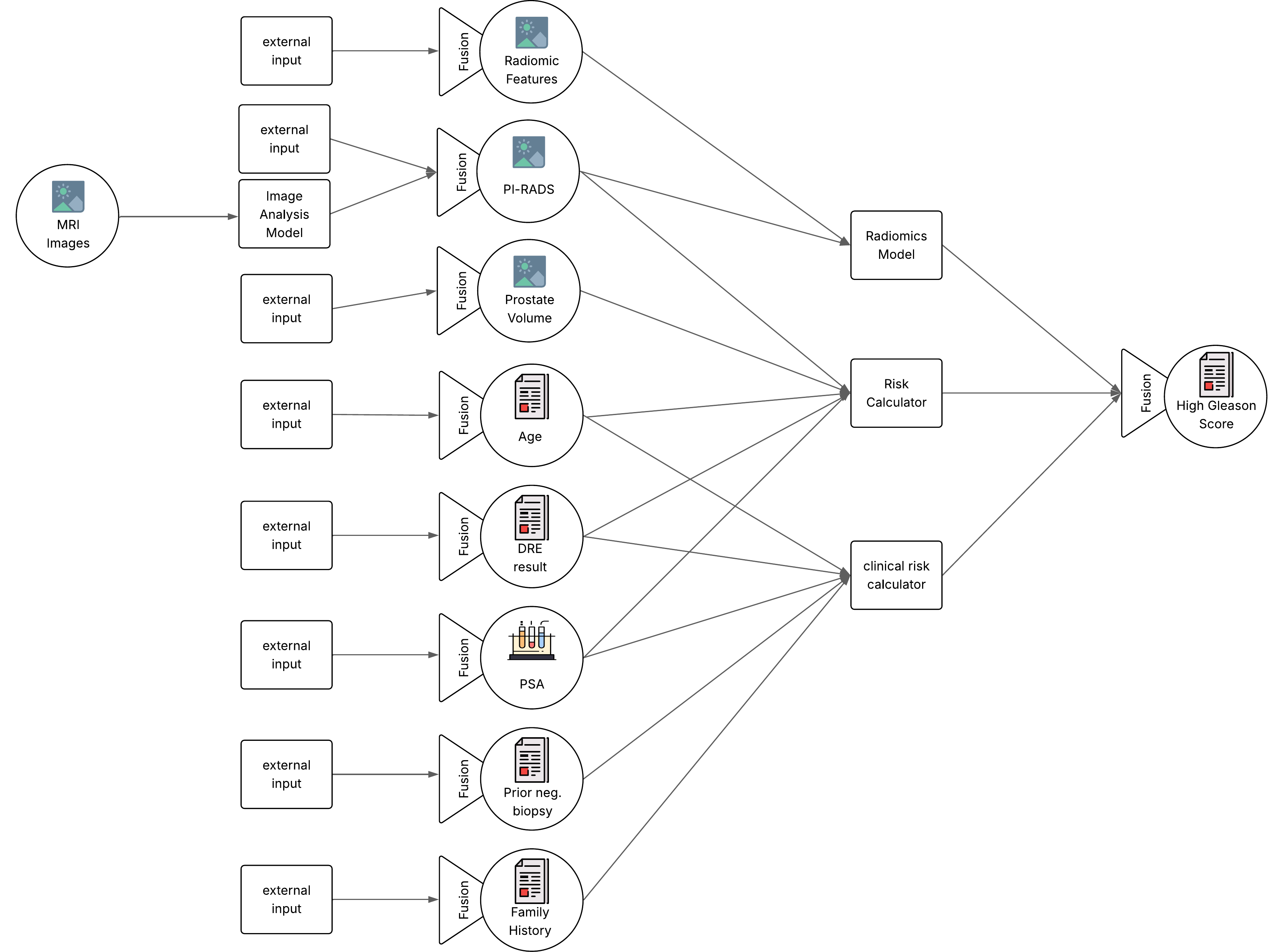}
    \caption{Detailed representation of the knowledge graph for "Patient John Smith" in Figure \ref{fig:ProstateFig2} before a prostate biopsy as constructed by the \textit{Backend Builder} (B$_3$). On the left, the input features are shown as circles, and the output attribute "High Gleason Score" is shown on the right. The models are depicted as squares. "external Input" models }
    \label{fig:ProstateGraph}
\end{figure*}

In Figure \ref{fig:ProstateFig2}, we have depicted this information flow as follows. Information about the patient is generally collected and governed within the \textit{clinical datasystems} (A) (Hospital Information System - HIS) in an unstructured and distributed way. For example, documentations of anamnesis might be in Word or PDF format. The \textit{data transformer} extracts and structures the patient information from each source and transfers it into the Digital Twin's \textit{data backbone} (B$_1$). An implementation of the \textit{data transformer} has to support a variety of widespread data protocols, as full interoperability in the healthcare context has not yet been reached \citesupplementary{TorabMiandoab2023}.
Features extracted from the anamnesis could be the family history regarding prostate cancer and other conditions, as well as the performance of prior biopsies. 
For our example, the data of "Patient John Smith" (including PSA value, DRE results and age) is stored in the \textit{data backbone} (B$_1$), together with the newly acquired MRI information that is additionally represented in the \textit{front-end} as \textit{input} (C$_1$). 

For the DT to mimic the information flow of the clinical patient journey, the \textit{attribute neighborhoods} within the \textit{Resource Description Framework} (B$_2$) have to be defined. Figure \ref{fig:ProstateFig2} shows two examples for \textit{attribute neighborhoods} that store the information on what informative base models enter the attribute-specific fusion model, as well as what base models are informed by the attribute. The first exemplified attribute is the PI-RADS score, for which two informative base models are entering the fusion model: the radiologist, represented as "external input" model, as well as the output of an AI-driven "Image Analysis" model - trained to predict the PI-RADS score based on the original MRI images. In such cases, where external and model-based inputs are available, the fusion is defined to select the external input ("overwrite mode"). 

Two base models are informed by the PI-RADS attribute, namely a radiomics model and a risk calculator, which will be explained in more detail later. The second exemplified attribute is the "high Gleason score", which corresponds to the next step in the clinical patient journey. Therefore, models that are defined as informed base models for the \textit{attribute neighborhood} of PIRADS are now part of the informative base models alongside a new machine learning based model. It is now clear how to combine both \textit{attribute neighborhoods} in the \textit{backend builder} (B$_3$). The PI-RADS will inform the radiomics model and the risk calculator will predict the high Gleason score attribute. 

Figure \ref{fig:ProstateGraph} visualizes the knowledge graph that is constructed upon the two introduced \textit{attribute neighborhoods} by the \textit{backend builder}. To simplify the example, we restrict the DT to three specific models, enough to show the main features of our operational mode. All of the models predict the probability of clinically significant prostate cancer. The first model is a model based purely on radiomics data, e.g., Jing et al. \citesupplementary{jingPredictionClinicallySignificant2022}. The second model is a mixed model, using clinical data and MRI information in the form of the PI-RAD score \citesupplementary{radtkeCombinedClinicalParameters2017a}. The last model is only using clinical and anamnesis data, i.e., age, psa value, DRE result, family history and whether a previous biopsy was negative. Such a model was developed, for example, by Ankerst et al. \citesupplementary{ankerstContemporaryProstateBiopsy2018}.

Before MRI information is available, only the clinical risk calculator has all the necessary data available. Thus, only this model propagates information to the high Gleason score attribute. The corresponding fusion receives information of this one model and does nothing more than adding the model's signature to the provenance chain (introduced in Figure 3 of the main) $\mathscr{P}$ of the high Gleason score attribute.

As the PI-RADS becomes available, the fusion model of the PI-RADS attribute notes the origin of the attribute PI-RADS in its corresponding provenance $\mathscr{P}$. As now a PI-RADS is available, the base models radiomics model and risk calculator check if all their inputs are available. In this example, this condition is fulfilled. 
Therefore, the models evaluate their inputs and provide an output (high Gleason score attribute). 
Additionally, according to the scheme in Fig. 4 of the main text, the models take the provenance chains of their inputs, combine them, add themselves. On the left side of Figure 4, this is depicted as $ \mathscr{P} \gets \{b\}$. The provenance chain gets passed together with the outputs to the fusion model. The corresponding fusion model detects updates of its inputs and as the input is not yet on the provenance chain of the high Gleason score attribute, the fusion model recalculates this attribute. This is done by a weighted average mechanism, using the previous performance of the models on similar patients as weights. Then it combines the provenance chains of all new inputs and adds them to the existing provenance chain of the attribute high Gleason score. On the right side of Figure 4 of the main text, this is included as $ \mathscr{P} \gets \{f\}$. This cycle then continues downstream, but for our example, we stop here at the first iteration. A fully implemented knowledge graph with this propagation and aggregation scheme would stop, either if no base model can be evaluated, because inputs are missing at this stage, or if a loop is detected, i.e. a model finding itself on the provenance chain. 

The calculated probability of the high Gleason score attribute will be presented to the clinician. This is the decision support. We expect the clinician to have the option to access all other attributes and their history, independent of whether they were calculated or measured values. Additionally, the provenance chain of the attribute enables tracking of the information flow, which can be visualized in a graph structure similar to Figure \ref{fig:ProstateGraph}.

The \textit{updater PD} (in B$_1$) stores the newly calculated information in the patient data. Once the patient has completed the patient journey (or just the current phase of the patient journey), the \textit{updater DC} stores the full patient information in the digital cohort. This time is different for every patient. The shortest period in our scenario would be the evaluation point of the biopsy, i.e., after a few days. At this time, the real value for the attribute is determined. After several patients have been added to the digital cohort, the \textit{updater RDF} updates the models in the RDF. The time interval depends on many factors, mainly hospital size and number of patients, but we can assume it to be in the order of once per year or less. 
Thus, for our "Patient John Smith", the doctor can now evaluate the predicted probability for a high Gleason score and, therefore, if he needs a biopsy or if it would be reasonable to further surveil PSA values.

\subsection{Survival Prediction in Glioblastoma}

\begin{figure*}
    \centering
    \includegraphics[width=\textwidth]{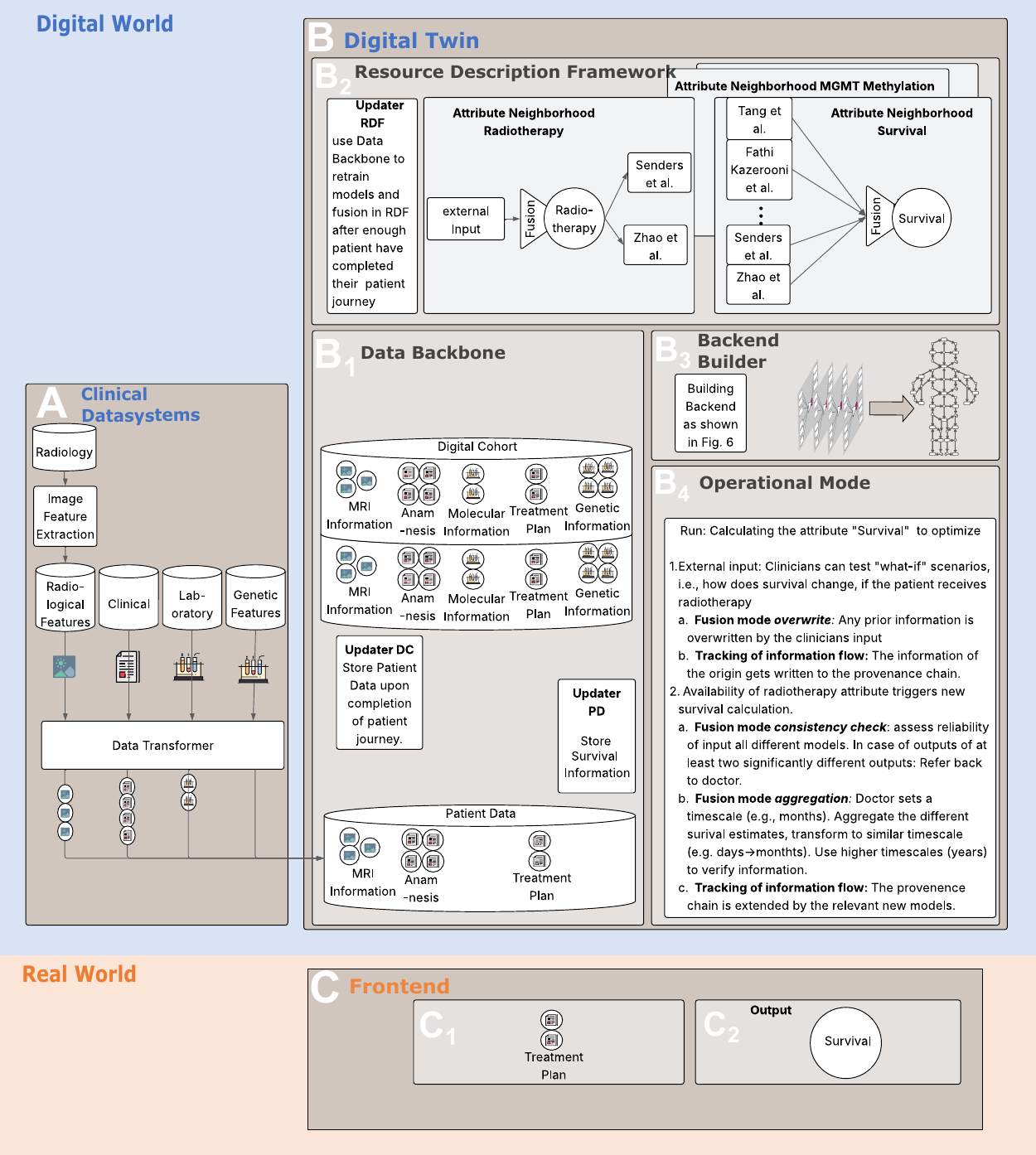}
    \caption{Schematic overview of our proposed software design for patient-centred DTs in glioblastoma survival estimation. The figure is similar to Fig. 2 of the main text, showing Clinical Datasystems (A), the DT itself (B), consisting of \textit{data backbone} (B$_1$), the \textit{Resource Description Framework} (B$_2$), which stores all available information about models and their links with attributes, which the \textit{back-end builder} (B$_3$) uses to construct a knowledge graph upon which the \textit{operational mode} (B$_4$) is performing predictions. The user interacts with the \textit{front-end} (C).}
    \label{fig:Glioma}
\end{figure*}

In this scenario, we are assuming a patient with glioblastoma multiforme (GBM), which is the most aggressive type of primary brain tumor. 
To support personalized treatment and avoid ineffective treatment, an accurate prediction of prognosis and survival is crucial \citesupplementary{poursaeedSurvivalPredictionGlioblastoma2024}. 

Traditionally, survival prognosis is based on a limited number of molecular features, e.g., MGMT methylation (methylation of the MGMT gene promoter reduces the production of the MGMT enzyme, which is a DNA repair protein that stronly reduce the patient responds to chemotherapy drugs like temozolomide), or clinical features, e.g., age and Karnofsky performance status (KPS describes the patient's physical state and abilities) \citesupplementary{wellerEANOGuidelinesDiagnosis2021}. 
Additionally, in recent years, a variety of machine learning based models for survival prediction have emerged, based on combinations of MR Images, radiomic, molecular, genomic, and clinical features \citesupplementary{poursaeedSurvivalPredictionGlioblastoma2024}. 
\begin{figure*}
    \centering
    \includegraphics[width=\linewidth]{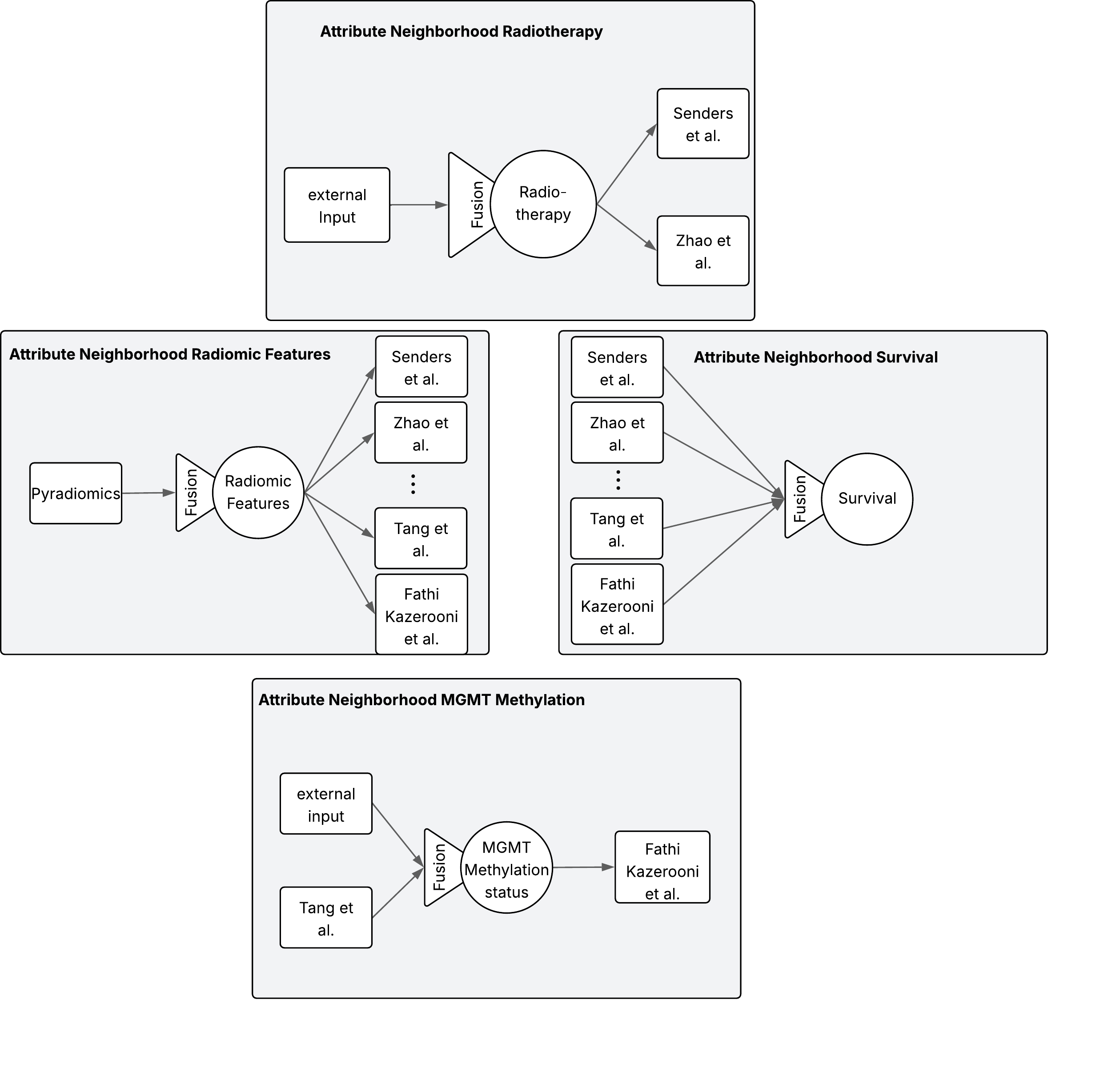}
    \caption{Selection of \textit{attribute neighborhoods} for survival prediction in glioblastomas. On top, we have the "Attribute Neighborhood Radiotherapy" attribute. This is the starting point. Radiotherapy information will be passed to two models, namely Senders et al. and Zhao et al. Both models are also in the "Attribute Neighborhood Radiomic Features". As we assume this information to be present, this feature can inform models of Tang et al. and Fathi Kazerooni et al. At the bottom, the "Attribute Neighborhood MGMT Methylation" shows, that both models, Tang et al. and Fathi Kazerooni et al. are connected via the MGMT methylation status. At the right side, the "Attribute Neighborhood Survival" is shown. It collects all models that estimate the survival of the patient. For clarity, only the four models also shown in the other attribute neighborhoods are visualized. The connected graph is then shown in Fig. \ref{fig:Glioma_Graph}}
    \label{fig:AttNei_GBM}
\end{figure*}

\begin{sidewaysfigure*}
    \centering
    \includegraphics[width=\textwidth]{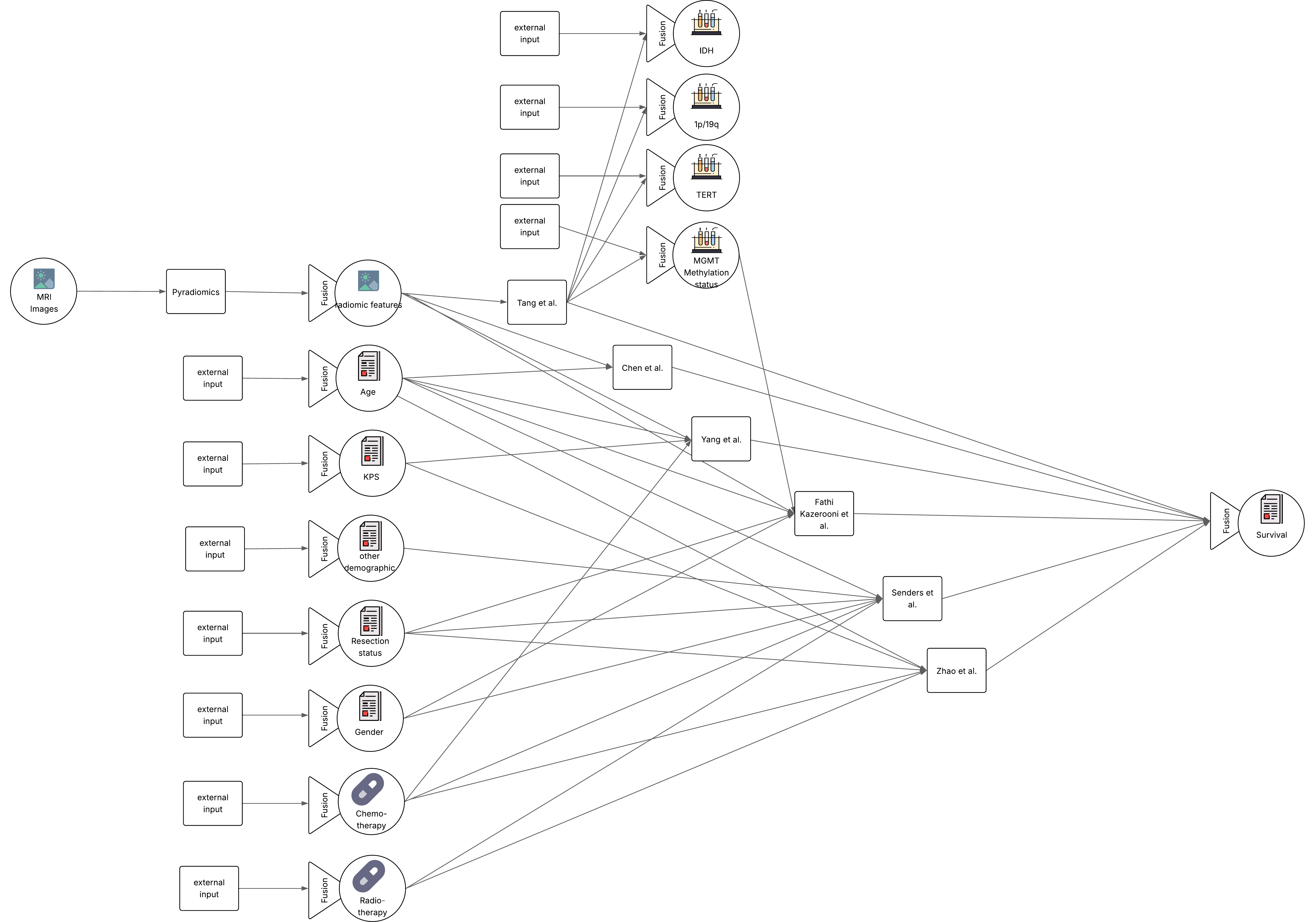}
    \caption{Detailed representation of the knowledge graph for "Patient Sarah Jane". The knowledge graph represents the interrelations of the models and data from imaging to survival prediction as
constructed by the Backend Builder (B3).}
    \label{fig:Glioma_Graph}
\end{sidewaysfigure*}

We now exemplify the DT for survival prediction, after the initial treatment for "Patient Sarah Jane", which was a surgical resection of the tumor. For the clinicians, the next decision is if therapy will be continued and, if so, if it will be chemotherapy or radiotherapy \citesupplementary{wellerEANOGuidelinesDiagnosis2021}. Therefore, models that estimate the survival based on the potential therapy types are needed. The information that has been collected to this point includes MRI scans, clinical data, and surgical outcomes, e.g., resection status. The evaluation metric would be a prolonged survival.

We explain the flow of data starting at the \textit{clinical datasystems} (A) in Figure \ref{fig:Glioma}. 
Similar to our previous example, we assume the \textit{data transformer} to transfer the information retrieved from the \textit{clinical datasystems} in a structured form in the \textit{data backbone} ($B_1$) of the Digital Twin (B). 
The data consists of imaging information, anamnesis, genetic features, molecular features, and potential previous treatments. After initial surgical resection of the tumor, the surgical margins are known ("resection status"), and the imaging procedure is repeated. Based on this information, the Digital Twin now calculates the survival estimate of our patient, without any further therapy. The doctor can input (C$_1$.2) a treatment option to simulate "what-if" scenarios for treatment. In the \textit{RDF} (B$_2$), the \textit{attribute neighborhood} of the radiotherapy is shown on the left side. It has only the external input, which can be set by the doctor, and informs two models. On the right side of the RDF (B$_2$), the \textit{attribute neighborhood} of the survival attribute is shown, with several base models informing this attribute. To emphasize the scalability of our approach, we have chosen a total of 6 base models, all with slightly different input and output features. A lot more models for predicting survival are available as reviewed by Poursaeed et al. \citesupplementary{poursaeedSurvivalPredictionGlioblastoma2024}.

The first model we chose is from Chen et al. and estimates the survival in days based on radiomic features and clinical features \citesupplementary{chenSubregionbasedSurvivalPrediction2023}. 
The second model was developed by Fathi Kazerooni et al. and categorizes patients into high risk (survival < 6 months), medium risk (survival 6-18 months), and low risk (survival > 18 months) based on radiomic features, molecular features, and clinical features \citesupplementary{fathikazerooniClinicalMeasuresRadiomics2022}. 
The third model, by Yang et al., predicts survival categories of 1 year, 2 years, or 3 years based on age, KPS, MGMT methylation status, and chemotherapy status \citesupplementary{yangSpatialHeterogeneityEdema2022}. 
The fourth model was developed by Tang et al. and predicts the survival in days, molecular features, in particular MGMT methylation, and genomic mutations. The prediction is based on a combination of MRI images, radiomic features, and clinical features \citesupplementary{tangDeepLearningImaging2020}. 
The fifth model is by Senders et al. and uses information on therapy, age, and resection status as inputs to predict a survival probability for 1 year, 2 years, or 3 years caategories \citesupplementary{sendersOnlineCalculatorPrediction2020}.
The sixth model by Zhao et al. predicts 6-month, or 12-month survival based on therapy form, resection status, KPS, and age \citesupplementary{zhaoOptimizingManagementElderly2022}.
Different from the example presented for the biopsy decision of prostate cancer, in this scenario, all base models output survival estimates, but vary with regards to their output details (days, months, years). The fusion model has the task of combining them in a meaningful way. A possible mechanism of the fusion model will be explained later. 

For the \textit{backend builder} (B$_3$) to build the \textit{knowledge graph}, it has to combine the most relevant attribute neighborhoods, which are visualized in Fig. \ref{fig:AttNei_GBM}, including the "Attribute Neighborhood Radiotherapy",  "Attribute Neighborhood Radiomic Features",  "Attribute Neighborhood MGMT Methylation" and  "Attribute Neighborhood Survival". As our patient has imaging information, we assume radiomic features are available for the Digital Twin. The radiomic features inform four models, namely Sanders et al., Zhao et al., Tang et al., and Fahi Kazerooni et al.. The model of Tang et al. not only predicts survival, but also genetic information from the MRI images, radiomic features, and clinical features. Specifically, the model predicts the MGMT methylation status, which, on the other hand, is an input for the model of Fathi Kazerooni et al.. 
Thus, the "MGMT methylation status" adds a connection between those two models in the knowledge graph. Although given such cross-connections, the complete graph will be more complex, the local structure will always be a simple coupling of different \textit{base models} via data attributes and their fusion models, which defines the \textit{attribute neighborhoods}. Figure \ref{fig:Glioma_Graph} visualizes the overall knowledge graph structure, including all additionally required patient attributes for the base models and their relations.
From left to right, we show the input data, then an intermediate layer, and finally our desired outcome attribute. The input layer is the available clinical data. The intermediate layer represents the cross-connection of the models of Tang et al. and Fathi Kazerooni et al., as both models are connected via an output-input relation, as explained before. At the far right end, all models inform the survival attribute and the corresponding fusion model.

The \textit{operational mode} controls the propagation and aggregation of information in the knowledge graph. Information from the MRI Images is extracted as radiomic features. The radiomic features are used by the models of Yang et al., Senders et al., and Zhao et al., which compute their survival estimate based on the given therapeutic approach. Additionally, Tang et al. use radiomic features together with the original images, age, and gender data to predict a survival estimate, MGMT methylation status and other mutation information. Although this model is independent of the treatment plan, we can envision such models existing. Such models can easily be added to the knowledge graph in the future.
The \textit{fusion model} for the MGMT methylation status has to combine a model output and an external input. Therefore, it is in an \textit{overwrite mode}, similar to our previous example for biopsy decision. If present, the fusion always prefers the external input and discards the model output. 

The remaining model of Fathi Kazerooni et al. has now all the needed input features and calculates its survival estimate. Now, all models have predicted the survival, but at different levels of detail (days, months, years). The fusion now has the task of combining all this information for the clinician into one best estimate of survival. This is the aggregation mode. Unlike a simple combination, the fusion has to transform all outputs so that they have the same meaning. For survival, this is the timescale. We assume this to be externally set by the doctor, e.g., he/she wants to know the chance for a 6-month survival. In this case, the fusion would use all models that provide the survival probability at this level of detail or higher level of detail, e.g., survival probability in days. The information with higher level of detail would be integrated to represent the wanted information, thus integrating from 0 to 180 days.  Any information with a lower level of detail, e.g., survival probability per year, would be used as verification, e.g., by ensuring the 1-year survival estimate should not exceed the 6-month survival estimate. If the verification fails, all information should be passed back to the doctor, and conflicts should be highlighted in an interactive output scheme. Then the information flow gets tracked by adding all models except the verification models to the provenance chain $\mathscr{P}$ of the survival attribute.

Finally, the \textit{updater PD} updates the Patient Data with the recent best estimates of survival. After the real survival of the patient is known, typically in the order of months to years, the \textit{updater DC} adds all measured information to the \textit{digital cohort}. After an even longer time period, the \textit{updater RDF} uses the past cases for retraining the individual base and fusion models in the knowledge graph. This timer period is in the order of years, but it depends on other factors, like the number of patients collected at the clinic and the heterogeneity of the clinical patient journey, and therefore, the availability and missingness of specific data.

By giving the best possible survival estimate for our patient dependent on a therapy, the clinician is now able to choose the best treatment plan.

\bibliographystylesupplementary{unsrt}
\bibliographysupplementary{zotero_static.bib} 

\end{document}